\documentclass[acmsmall,screen]{acmart}
\usepackage[utf8]{inputenc}
\usepackage{microtype}
\usepackage{wrapfig}
\usepackage{listings}
\usepackage{xspace}
\usepackage{mathpartir}
\usepackage{subcaption} 
\usepackage[inference]{semantic}
\usepackage{booktabs}   
\usepackage{listings}
\usepackage{array}
\usepackage{multirow}
\usepackage{graphicx}
\usepackage{caption}
\usepackage{makecell}
\usepackage{algorithm}
\usepackage{algpseudocode}
\usepackage{soul}
\usepackage{float}
\newfloat{algorithm}{t}{lop}

\AtBeginDocument{%
  }

\setcopyright{rightsretained}
\acmPrice{}
\acmDOI{10.1145/3591285}
\acmYear{2023}
\copyrightyear{2023}
\acmSubmissionID{pldi23main-p414-p}
\acmJournal{PACMPL}
\acmVolume{7}
\acmNumber{PLDI}
\acmArticle{171}
\acmMonth{6}
\received{2022-11-10}
\received[accepted]{2023-03-31}

\def\sng#1{\textcolor{blue}{\sf SNG:$\clubsuit$ #1$\clubsuit$}}

\newcommand\absynthe{\textsc{Absynthe}\xspace}
\newcommand\sygus{SyGuS\xspace}
\newcommand\autopandas{\textsc{AutoPandas}\xspace}

\newcommand\rbsyn{\textsc{RbSyn}\xspace}

\def\spmid{\ \mid \ }

\newcommand\strlang{\ensuremath{\mathcal{L}_{str}}\xspace}
\newcommand\flang{\ensuremath{\mathcal{L}_{f}}\xspace}
\newcommand\metalang{\ensuremath{\mathcal{L}_{meta}}\xspace}

\newcommand\judgementHead[1]{\hfill\fbox{\ensuremath{#1}}}

\newcommand\vtrue{\texttt{true}\xspace}
\newcommand\vfalse{\texttt{false}\xspace}

\newcommand{\powerset}{\raisebox{.15\baselineskip}{\large\ensuremath{\wp}}\xspace}

\newcommand\valset{\ensuremath{\mathbb{V}}\xspace}
\newcommand\absconst{\ensuremath{\tilde{\val}}\xspace}
\newcommand\absvar{\ensuremath{\tilde{\var}}\xspace}
\newcommand\absval{\ensuremath{a}\xspace}
\newcommand\absvalset{\ensuremath{\mathbb{A}}\xspace}
\newcommand\func[2]{\ensuremath{f(#1, \ldots, #2)}\xspace}
\newcommand\absfunc[2]{\ensuremath{f^\#(#1, \ldots, #2)}\xspace}

\newcommand\strlenval{\ensuremath{l}\xspace}

\newcommand\branch{\ensuremath{b}\xspace}
\newcommand\val{\ensuremath{v}\xspace}
\newcommand\expr{\ensuremath{e}\xspace}
\newcommand\var{\ensuremath{x}\xspace}

\newcommand\eif[3]{\ensuremath{\textbf{\texttt{if}}\ #1\ \textbf{\texttt{then}}\ #2\ \textbf{\texttt{else}}\ #3}\xspace}

\newcommand\eprog[3]{\ensuremath{\textbf{\texttt{def}}\ #1(#2) = #3}}

\newcommand\ehole{\ensuremath\square\xspace}

\newcommand\mthtype[2]{\ensuremath{#1 \rightarrow #2}\xspace}

\newcommand\rulename[1]{\textsc{#1}\xspace}

\newcommand\tenv{\ensuremath{\Gamma}\xspace}
\newcommand\benv{\ensuremath{\Delta}\xspace}

\newcommand\meth{\ensuremath{m}\xspace}
\newcommand\program{\ensuremath{P}\xspace}

\newcommand\ie{\textit{i.e.,}\xspace}
\newcommand\eg{\textit{e.g.,}\xspace}
\newcommand\btime[2]{#1 \scriptsize{\ensuremath{\pm} #2}}
\newcommand\emptyspace{\ensuremath{``\ \ "}}

\newcommand\metafun[1]{\ensuremath{g(#1)\xspace}}

\newcommand\semgus{\textsc{SemGuS}\xspace}
\newcommand\dslvar{\ensuremath{y}\xspace}
\newcommand\dslexpr{\ensuremath{\hat{\expr}}\xspace}

\definecolor{greenhl}{rgb}{0.40,0.72,1.00}
\definecolor{redhl}{rgb}{1.00,0.69,0.00}
\DeclareRobustCommand{\hlgreen}[1]{{\sethlcolor{greenhl}\hl{#1}}}
\DeclareRobustCommand{\hlred}[1]{{\sethlcolor{redhl}\hl{#1}}}

\definecolor{programs}{gray}{0.1}
\definecolor{keywords}{HTML}{204a87}
\definecolor{comments}{HTML}{8f5902}
\definecolor{strings}{HTML}{4e9a06}

\lstnewenvironment{rcodebox}
{\lstset{upquote=true,xleftmargin=2em,language=Ruby,breaklines=true,numbers=left,stepnumber=1,firstnumber=1,numberfirstline=true, frame=none, numberstyle={\footnotesize\it\color{programs}}}
}
{}
\newcommand\rcode{\lstinline[language=Ruby,mathescape,basicstyle=\sffamily\normalsize,breakatwhitespace,xleftmargin=0pt,xrightmargin=0pt]}

\begin{document}

\title{\textsc{Absynthe}: Abstract Interpretation-Guided Synthesis}

\author{Sankha Narayan Guria}
\affiliation{
  \institution{University of Maryland}            
  \city{College Park}
  \state{Maryland}
  \postcode{20742}
  \country{USA}                    
}
\email{sankha@cs.umd.edu}          

\author{Jeffrey S. Foster}
\affiliation{
  \institution{Tufts University}            
  \city{Medford}
  \state{Massachusetts}
  \postcode{20742}
  \country{USA}                    
}
\email{jfoster@cs.tufts.edu}          

\author{David Van Horn}
\affiliation{
  \institution{University of Maryland}            
  \city{College Park}
  \state{Maryland}
  \postcode{20742}
  \country{USA}                    
}
\email{dvanhorn@cs.umd.edu}          





\begin{abstract}
Synthesis tools have seen significant success in recent times. However,
past approaches often require a complete and accurate embedding of the source
language in the logic of the underlying solver, an approach difficult for
industrial-grade languages. Other approaches couple the semantics of the source
language with purpose-built synthesizers, necessarily tying the synthesis engine
to a particular language model.
In this paper, we propose \absynthe, an alternative approach based on
user-defined abstract semantics that aims to be both lightweight and language
agnostic, yet effective in guiding the search for programs.
A synthesis goal in \absynthe is specified as an abstract
specification in a lightweight user-defined abstract domain and concrete test
cases.
The synthesis engine is parameterized by the abstract semantics and independent
of the source language.
\absynthe validates candidate programs against test cases using the
actual concrete language implementation to ensure correctness.
We formalize the synthesis rules for \absynthe
and describe how the key ideas are scaled-up in our implementation in Ruby. We
evaluated \absynthe on \sygus strings benchmark and found it competitive with
other enumerative search solvers. Moreover, \absynthe's ability to combine
abstract domains allows the user to move along a cost spectrum, \ie expressive
domains prune more programs but require more time. Finally, to verify
\absynthe can act as a general purpose synthesis tool, we use \absynthe to
synthesize Pandas data frame manipulating programs in Python using simple
abstractions like types and column labels of a data frame. \absynthe reaches
parity with \autopandas, a deep learning based tool for the same benchmark
suite. In summary, our results demonstrate \absynthe is a
promising step forward towards a general-purpose approach to synthesis that may
broaden the applicability of synthesis to more full-featured languages.

\end{abstract}



\maketitle

\section{Introduction}

In recent years, there has been a significant interest in automatically
synthesizing programs from high-level specifications, which often take the form
of logical formulas~\cite{FengMBD18},
type signatures~\cite{Polikarpova2016Synquid},
or even input-output examples~\cite{frankle2016example}. Program synthesis has
seen significant
success in domains such as spreadsheets~\cite{Gulwani11}, compilers~
\cite{PhothilimthanaE19} or even database access programs~\cite{rbsyn-pldi21}.
Much of the prior work, however, requires a complete and accurate embedding of
the source language in the logic of the underlying solver the synthesis tool
uses. These often range from symbolic execution~\cite{TorlakRosette14},
counter-example guided synthesis~\cite{SolarLezamaSketch13}, or over-approximate
semantics as predicates~\cite{KimSemGuS21,Polikarpova2016Synquid,fengtable2017} 
(often requiring termination measures and additional predicates for verifcation).
This is infeasible for many industrial-grade languages such as Ruby or Python.
Other approaches are strongly coupled with the semantics of the source language
with purpose-built solvers~\cite{ReynoldsDKTB15}, but this necessarily ties the
synthesis engine to the particular language model used.


In this paper, we propose \absynthe, an alternative approach based on
user-defined abstract semantics that aims to be both lightweight and language
agnostic. The abstract semantics are lightweight to design, simplifying away
inconsequential language details, yet effective in guiding the search for
programs. The synthesis engine is parameterized by the abstract
semantics~\cite{CousotC77} and independent of the source language.
In \absynthe, users define a synthesis problem
via concrete test cases and an abstract specification in some
user-defined abstract domain. These abstract domains, and the semantics of the
target language in terms of the abstract domains, are written by the user in a
DSL. Moreover, the user can define multiple simple domains, each defining a
partial semantics of the language, which they can combine together as a product
domain automatically. \absynthe uses these
abstract specifications to automatically guide the search for the program using
the abstract semantics. The key novelty of \absynthe is that it separates the
search procedure from the definition of abstract domains, allowing the search to
be guided by any user-defined domain that fits the synthesis task.
More specifically, the program search in \absynthe begins with a hole tagged
with an
abstract value representing the method's expected return value.
At each step, \absynthe
substitutes this hole with expressions, potentially containing more holes, until
it builds a concrete expression without any holes.
Each concrete expression generated is finally tested in the reference
interpreter to check if
it passes all test cases. A program that passes all tests
is considered the solution. (\S~\ref{sec:overview} gives a complete
example of \absynthe's synthesis strategies).

We formalize \absynthe for a core language \flang and define
an abstract interpreter for \flang in terms of abstract transformer functions.
Next, we describe a DSL \metalang used to define these abstract transformers.
Notably, as \absynthe synthesizes terms at each step, it creates
holes tagged as abstract variables \absvar, \ie holes which will be assigned a
fixed abstract values later.
We give evaluation rules for these transformers
written in \metalang, that additionally narrows these abstract variables to
sound range of abstract values.
For example, given a specification that requests
Pandas programs that should evaluate to a data frame, a term 
\rcode{($\ehole: \absvar_1$).query($\ehole: \absvar_2$)} is a viable
candidate that queries a data frame. However, semantics of \metalang help with
constraining the bounds on $\absvar_1$ and $\absvar_2$ such that these holes are
substituted by values of a \rcode{DataFrame} and \rcode{String} respectively.
Finally, we present the
synthesis rules used by \absynthe to generate such terms. 
Specifically, we discuss how \absynthe specializes term generation based on the
properties of the domain, such as a finite domain enables enumeration through
domain, or a domain that can be lifted to solvers can use solver-backed
operations, or domains expressed as computations not supported
by dedicated SMT solvers. (\S~\ref{sec:formalism} discusses our formalism).

We implemented \absynthe as a core library in Ruby, that provides the necessary
supporting classes to implement user-specific abstract domains and abstract
interpretation semantics. It further integrates automatic support for
$\top$ and $\bot$ values and abstract variables , as well as 
\rcode{ProductDomain} to combine the individual domains
point-wise.
The \absynthe implementation has interfaces to call a
concrete interpreter with a generated program to check if a program satisifies
the input/output examples. Finally, we also discuss some optimizations to scale
\absynthe for practical problems, such as caching small terms and guessing
partial programs based on testing predicates on the input/output
examples, and
some limitations of the tool.
(\S~\ref{sec:implementation} discusses our implementation).

We evaluate \absynthe as a general-purpose tool on a diverse set of
synthesis problems while being at par on performance with state-of-the-art
tools. We first use \absynthe to solve the \sygus strings benchmarks~
\cite{alur2017sygus} using simple domains such as string
prefix, string suffix, and string length to guide the search.
Though \absynthe operates with minimal semantic information about \sygus
programs, it still performs similar to enumerative search solvers
such as \textsc{EuPhony}~\cite{alureusolver2017}, solving most benchmarks in
less than 7 seconds. 
SMT solvers such as CVC4, or \textsc{Blaze} that rely on precise
abstractions perform much faster than \absynthe, but require
large specification effort.
We further evaluate the impact of our performance optimizations
and verify that \absynthe's synthesis cost adjusts with the
expressiveness of
the domain. More specifically, the string prefix and suffix domains
written in Ruby generate a concrete candidate 0.41ms average, whereas
string length domain being a solver-aided domain takes around 16.93ms per
concrete candidate on average due to calls to Z3.
Next, we use \absynthe synthesize an unrelated benchmark suite, for which it is
harder to write precise formal semantics---Python programs that use Pandas,
a data frame manipulation library. We evaluate \absynthe on the \autopandas
benchmarks~\cite{bavishi2019autopandas}, a suite of Pandas data frame
manipulating programs in Python. The \autopandas tool trains deep neural
network models to synthesize Pandas programs.
\absynthe is at par with \autopandas, including a
significant overlap in the benchmarks both tools can solve, despite using simple
semantics such types and column labels of a data frame while running on a
consumer Macbook Pro without specialized hardware requirements.
(\S~\ref{sec:evaluation} discusses our evaluation).


In summary, we think \absynthe represents an important step forward in the
design of practical synthesis tools that provide lightweight formal guarantees
while ensuring correctness from tests.

\section{Overview}
\label{sec:overview}

In this section, we demonstrate \absynthe{} by using it to synthesize data frame
manipulation programs in Python using the Pandas
library~\cite{pandas144}. In this example, we abstract
data frames as sets of column names, and use a lightweight type system for
Pandas API methods to effectively guide synthesis.



\begin{figure}[tbp]
  \begin{minipage}{0.6\linewidth}
  \centering
  \begin{subfigure}[b]{0.45\textwidth}
    \centering
    \begin{tabular}{|l|l|l|}
    \hline
      & \thead{id}  & \thead{valueA} \\
    \hline
    0 & 255 & 1141   \\
    1 & 91  & 1130   \\
    2 & 347 & 830    \\
    \vdots & \vdots & \vdots \\
    8 & 225 & 638    \\
    9 & 257 & 616    \\
    \hline
    \end{tabular}
    \caption{First data frame (\rcode{arg0})}
    \label{table:input1}
  \end{subfigure}
  \hfil
  \begin{subfigure}[b]{0.45\textwidth}
    \centering
    \begin{tabular}{|l|l|l|}
      \hline
         & \thead{id}   & \thead{valueB} \\
      \hline
      0  & 255  & 1231   \\
      1  & 91   & 1170   \\
      2  & 5247 & 954    \\
      \vdots & \vdots & \vdots \\
      12 & 211  & 575    \\
      13 & 25   & 530    \\
      \hline
      \end{tabular}
    \caption{Second data frame (\rcode{arg1})}
    \label{table:input2}
  \end{subfigure}

  \begin{subfigure}[b]{\textwidth}
    \vspace{2ex}
    \centering
      \begin{rcodebox}
def goal(arg0, arg1, arg2):
  return arg0.merge(arg1, on=['id']).query(arg2)
      \end{rcodebox}
    \caption{Synthesized program}
    \label{fig:synthesized-prog}
  \end{subfigure}

\end{minipage}
\hfil
\begin{minipage}{0.3\linewidth}
\begin{subfigure}[b]{\textwidth}
\begin{tabular}{|l|l|l|l|}
  \hline
  & \thead{id}  & \thead{valueA} & \thead{valueB} \\
\hline
0 & 255 & 1141   & 1231   \\
1 & 91  & 1130   & 1170   \\
2 & 347 & 830    & 870    \\
5 & 159 & 715    & 734    \\
8 & 225 & 638    & 644    \\
\hline
\end{tabular}
\caption{Output data frame}
\label{table:output}
\end{subfigure}

\begin{subfigure}[b]{\textwidth}
  \vspace{5ex}
  \centering
    \begin{rcodebox}
'valueA != valueB'
    \end{rcodebox}
    \caption{Query string (\rcode{arg2})}
    \label{fig:input3}
\end{subfigure}
\end{minipage}
  \caption{Data Frame Manipulation Example~\cite{bavishi2019autopandas}. The synthesis goal is to produce~(\subref{table:output}) given inputs~(\subref{table:input1}), ~(\subref{table:input2}), and a query string ~(\subref{fig:input3}). \absynthe synthesizes the solution (\subref{fig:synthesized-prog}).}
  \label{table:overview-example-pandas}
\end{figure}

A \emph{data frame} is a collection of data organized into rows and columns,
similarly to a database table. Data frame manipulation is a key task in data
wrangling, a preliminary step for data science or scientific computing tasks.
For example, Figure~\ref{table:overview-example-pandas} shows a data
frame manipulation synthesis task taken from the \autopandas benchmark
suite~\cite{bavishi2019autopandas}. The goal is to use the Python
Pandas library~\cite{pandas144} to produce the data frame in
Figure~\ref{table:output}, given the two input data frames in Figure~
\ref{table:input1} and~\ref{table:input2} and a query string (Figure~
\ref{fig:input3}). In this case, the output joins
the input rows with the same \rcode{id} but with different values in
\rcode{valueA} and \rcode{valueB} columns.
The Pandas library provides a wide range of methods that perform complex
data frame manipulation. For example, calling \rcode{left.merge(right, on: ['col'])}
joins the data frames \rcode{left} and \rcode{right} on column \rcode{col}. As
another example, calling \rcode{df.query(str)} returns a new data frame with the
rows of \rcode{df} that satisfy query string predicate \rcode{str} (as in
Figure~\ref{fig:input3}).

To keep the synthesis task tractable, \absynthe restricts its search to Python
code consisting of input variables \rcode{arg0} through \rcode{arg2}; constants
such as column names \rcode{'id'}, \rcode{'valueA'}, and \rcode{'valueB'} or row
labels 0, 1, $\ldots$, 13 from the data frames; array literals and indexing;
and dictionaries (for keyword arguments). Additionally, for this discussion we will
limit \absynthe to the \rcode{merge} and \rcode{query} methods just mentioned,
even though our evaluation (\S~\ref{subsec:eval:autopandas}) supports many more
methods. Nonetheless, even with this restricted search space, na\"ive
enumeration of possible solutions times out after 20 minutes. In contrast, using
\absynthe{}, we can guide the search using abstract interpretation to find a
solution in 0.47 seconds.

\begin{figure}
\begin{subfigure}[t]{0.45\linewidth}
\begin{rcodebox}
class ColNames < AbstractDomain
  attr_reader :cols (*\label{line:attr_reader_cols}*)
  def initialize(cols)
    @cols = cols.to_set
  end

  def (*$\subseteq$*)(rhs)
    rhs.cols.subset?(@cols)
  end

  def (*$\cup$*)(rhs)
    ColNames.new(@cols (*$\cup$*) rhs.cols)
end end
\end{rcodebox}
\caption{ColNames Domain}
\label{fig:colnames-defn}
\end{subfigure}
\hfil
\begin{subfigure}[t]{0.45\linewidth}
\begin{rcodebox}
class ColNameInterp < AbsInterp
  def self.interpret(env, prog)
    # details omitted for brevity
  end

  def self.pd_merge(left,right,opt)
    left (*$\cup$*) right (*\label{line:eval_pd_merge}*)
  end

  def self.pd_query(df, pred)
    df (*\label{line:eval_pd_query}*)
  end
end
\end{rcodebox}
\caption{ColNames Abstract Semantics}
\label{fig:colnames-semantics}
\end{subfigure}

\bigskip{}

\begin{subfigure}[t]{0.45\linewidth}
\begin{rcodebox}
class PyType < AbstractDomain
  attr_reader :ty

  def initialize(ty)
    @ty = ty
  end

  def (*$\subseteq$*)(rhs)
    @ty <= rhs.ty
  end
end
\end{rcodebox}
\caption{PyType Domain}
\label{fig:pytype-defn}
\end{subfigure}
\hfil
\begin{subfigure}[t]{0.45\linewidth}
\begin{rcodebox}
class PyTypeInterp < AbsInterp
  def self.pd_merge(left,right,opt)
    DataFrame if left (*$\subseteq$*) DataFrame &&
          right (*$\subseteq$*) DataFrame &&
          opt (*$\subseteq$*) {on: Array<String>}
  end

  def self.pd_query(df, pred)
    DataFrame if df (*$\subseteq$*) DataFrame &&
          pred (*$\subseteq$*) String
end end
\end{rcodebox}
\caption{PyType Abstract Semantics}
\label{fig:pytype-semantics}
\end{subfigure}
\caption{Abstract domain definition for column names domain~(\subref{fig:colnames-defn}) and types domain definition~(\subref{fig:pytype-defn}). Abstract semantics for the required methods are defined in~(\subref{fig:colnames-semantics}) using ColNames domain and~(\subref{fig:pytype-semantics}) using PyType domain.}
\label{fig:overview-domains-defn}
\end{figure}

\paragraph{Abstract Domains and Semantics}

The first step in using \absynthe{} is to identify appropriate abstract domains
for the abstract interpretation and implement the abstract semantics. Typically,
we develop the domain by looking at the input/output examples and thinking about
the problem domain. In our running example, we observe that the data frames use
columns \rcode{id}, \rcode{valueA}, and \rcode{valueB}, but each frame has a
slightly different set of columns. This gives us the idea of introducing a
domain \rcode{ColNames} that abstracts data frames to a set of column labels.

Abstract values, drawn from an abstract domain, represent a set of concrete
values in the program. The abstract semantics define the evaluation rules
of the program under values from this abstract domain. This approach has seen
considerable success in practical static analysis tools such as
ASTRE\'E~\cite{CousotCFMMMR05} and Sparrow~\cite{OhHLLY12}.
Figure~\ref{fig:colnames-defn} shows, similar to these tools, the definition of
the \rcode{ColNames} domain, which is a class whose instances are domain values.
\absynthe{} is
implemented in Ruby, and \absynthe{} domains subclass \rcode{AbstractDomain},
which provides foundational definitions such as $\top$ and $\bot$ (see
\S~\ref{sec:implementation}). A value in the \rcode{ColNames} domain stores the set
of columns it represents in the instance variable \rcode{@cols}, which by line~
\ref{line:attr_reader_cols} can be read with an accessor method \rcode{cols}.
All abstract domains \emph{require} a partial ordering relation $\subseteq$ on
the domain that returns true if and only if the first columns label set (
\rcode{rhs.cols}) is a subset of the second set (\rcode{@cols}).
Finally, the $\cup$ method returns a new abstract value containing the
union of the column names of the two arguments. The $\cup$ method is optional,
however we define this as it will be used in the abstract semantics.


After defining the abstract domain, next we need to define the abstract
interpreter to give semantics to the target language in our abstract domain.
Figure~\ref{fig:colnames-semantics} defines the abstract interpreter for 
\rcode{ColNames} domain as \rcode{ColNameInterp} class. All abstract
interpreters are defined as a subclass of \rcode{AbsInterp} class,
provided by \absynthe. It needs a definition of the \rcode{interpret} class
method (the preceding \rcode{self.} denotes it is a class method), that given
an environment \rcode{env}, and a term \rcode{prog} reduces it to a value of
type \rcode{ColNames}. The \rcode{interpret} is a standard recursive
interpreter, so we omit the definition of \rcode{interpret} for brevity.
Then we define the \rcode{pd_merge} and \rcode{pd_query} class methods
that define the operations for the Pandas \rcode{merge} and \rcode{query}
methods on values from \rcode{ColNames} domain.
A call to \rcode{left.merge(right, opt)} in the source term under abstract
interpretation is computed via a call
\rcode{pd_merge(abs(left), abs(right), abs(opt))}, where
\rcode{abs()} indicates the abstract values of the arguments.
In the column name abstraction, we only need to compute the column names of
the resulting data frame, which is just the union of the column names of the
input data frame (line~\ref{line:eval_pd_merge}). Notice the \rcode{opt}
argument can be ignored, as it impacts how the data frames are merged in the
concrete domain, but the set column names of the final data frame is unaffected.
Similarly, a call to \rcode{df.query(pred)} is abstractly evaluated via a call
to \rcode{pd_query(abs(df), abs(pred))}. Since the data frame returned by 
\rcode{query} has the same columns as its input data frame,
\rcode{pd_query} simple returns the abstract data frame \rcode{df}
(line~\ref{line:eval_pd_query}).

\absynthe{} can also combine multiple domains together pointwise. We observe
that the Pandas API methods expect values of a specific type. Hence, we also
introduce a \rcode{PyType} abstract domain as a lightweight type system for
Python. Figure~\ref{fig:pytype-defn} defines the abstract domain, which stores a
type in the \rcode{@ty} field as a type from RDL~\cite{rdl-github}, a Ruby type
system. We build on RDL for representing
Python types because it comes with built-in representations for nominal types,
generic types, etc. and a subtyping relation between them. The $\subseteq$
method for \rcode{PyType} simply calls the subtyping method $\leq$ of RDL types.
The subtyping method $\leq$ is a special-case of the partial ordering relation
$\subseteq$.

Figure~\ref{fig:pytype-semantics} defines gives the abstract semantics for 
\rcode{merge} and \rcode{query} in the \rcode{PyType} domain. The method
\rcode{pd_merge} checks that the types of \rcode{left} and \rcode{right} are
subtypes of \rcode{DataFrame}, \ie the type that represents Pandas data frames
as shown in Figure~\ref{table:overview-example-pandas}, and that 
\rcode{opt} is a dictionary with a key \rcode{on} that admits an array of
strings. If this check is satisfied, the return type is \rcode{DataFrame}.
Otherwise, \rcode{pd_merge} returns \rcode{nil}, which \absynthe interprets
as $\top$, \ie any value is possible. 
Note, in a type checker, if the arguments do not match the expected types a type
error occurs. Here, in contrast, we are computing what would be a valid
abstraction, and since we don't have a specific type we can assume $\top$, \ie
anything can happen. Later, during synthesis the search procedure will
appropriately do the pruning by \emph{type-checking} when it is provided a user
specification.
\rcode{pd_query} also checks if the receiver is a subtype of
\rcode{DataFrame} and the query string is a \rcode{String}. If so, it returns
\rcode{DataFrame}, otherwise it returns \rcode{nil}.

These domains are combined together using a \rcode{ProductDomain} class,
provided by \absynthe. Here we write $\times$ to pair elements from the 
\rcode{ColNames} domain and the \rcode{PyType} domain.
For example, $\{\textnormal{\rcode{'id', 'valueA'}}\} \times
\textnormal{\rcode{DataFrame}}$ denotes all data frames that have the columns
\rcode{'id'} and \rcode{'valueA'}. The \rcode{ProductDomain} also comes with a
\rcode{ProductInterp} that evaluates product domain values with respective
individual semantics and combines these into a final product abstract value.

\paragraph{Synthesizing Solutions}

\begin{figure}
  \centering
  \includegraphics[scale=0.41]{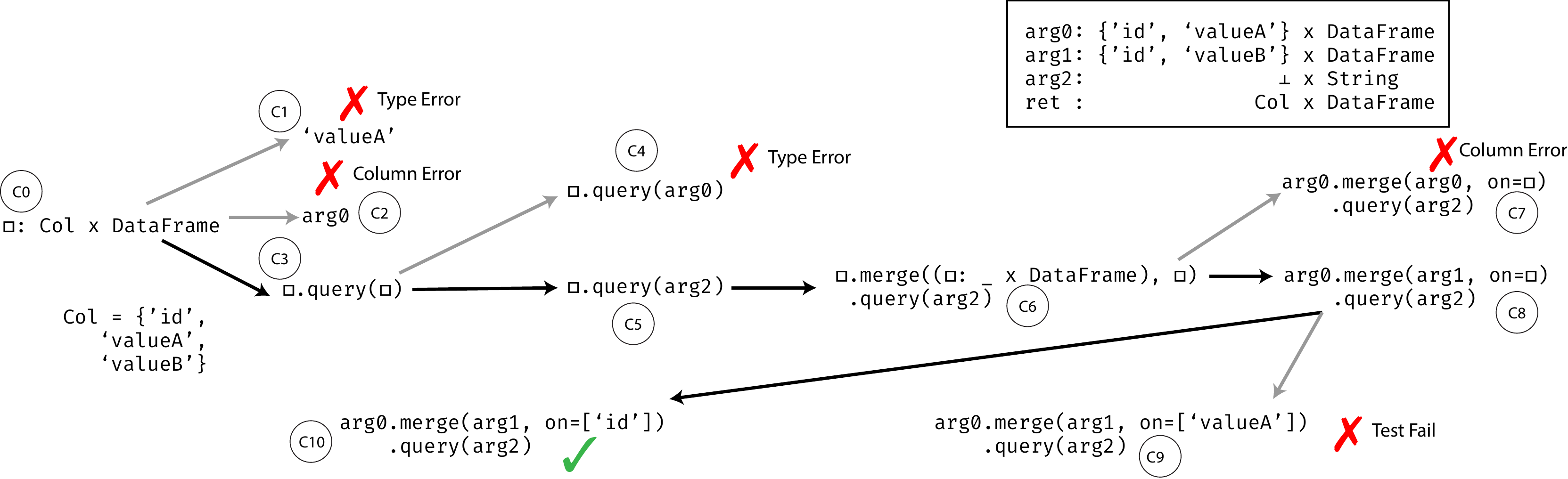}
  \caption{Steps in the synthesis of solution to the problem in Figure~
  \ref{table:overview-example-pandas}. Some choices available to the synthesis
  algorithm has been omitted for simplicity.}
  \label{fig:synthesis-flow}
\end{figure}

An \absynthe{} synthesis problem is specified by giving input/output examples
for the synthesized function. Synthesis begins by abstractly interpreting the
input/output examples to compute an \emph{abstract signature} for the function.
We have automated this for the \autopandas benchmark suite.
The upper-right corner of Figure~\ref{fig:synthesis-flow} gives  the abstract
signature for out example. In particular, the first argument is a 
\rcode{DataFrame} with columns \rcode{'id'} and \rcode{'valueA'}; the second
argument is a \rcode{DataFrame} with columns \rcode{'id'} and \rcode{'valueB'};
and the third argument is a \rcode{String} and has no columns.
The synthesized function should return a  \rcode{DataFrame} that has columns 
\rcode{'id'}, \rcode{'valueA'}, and \rcode{'valueB'}.
Additionally, \absynthe also uses a set of constants that can be used during the
synthesis process. It constructs this from the rows and columns of the
dataframes in the input/output example: $\{$\rcode{'id'}, \rcode{'valueA'},
\rcode{'valueB'}$, 0, 1, \ldots, 13 \}$.

\absynthe{} iteratively produces candidate function bodies that may contain 
\emph{holes} $\ehole: a$, where each hole is labeled with the abstract value $a$
its solution must abstractly evaluate to. Synthesis begins (left side of figure)
with candidate \textsf{C0}, which is a hole labeled with the abstract return
value of the function.
At each step, \absynthe replaces a hole with an expression that satisfies its
labels. For example, candidate \textsf{C1} is not actually generated because its
concrete value \rcode{'valueA'} is not of type \rcode{DataFrame}.
The process continues until the program has been full concretized, at which
point is it tested in the Python interpreter against the input/output examples.
Synthesis terminates when it finds a candidate that matches the input/output
examples. For our running example, Figure~\ref{fig:synthesized-prog} shows the
solution synthesized by \absynthe.

The rest of the figure illustrates the search process.
The candidate \textsf{C2} does not satify the abstract specification on columns,
so it is also never generated.
The candidate \textsf{C3} instead expands the hole to a call to \rcode{query},
which itself has holes for the receiver and argument.
Note we omit the abstraction labels here because \absynthe has not fixed the
abstract value for that hole yet.
\absynthe treats these as abstract variables that can be used during abstract
interpretation, but will be eventually substituted with a fixed abstract value
as the search proceeds (discussed in \S~\ref{sec:formalism}).

After a single set of expansion of holes, \absynthe runs the abstract
interpreter on all candidates (including the partial programs). Running 
\textsf{C3} through the abstract interpreter calls the \rcode{pd_query} function
from \rcode{PyTypeInterp} (Figure~\ref{fig:pytype-semantics}). From the
evaluation of the \rcode{pd_query} \absynthe can infer the first hole has to be
a subtype of \rcode{DataFrame} and the second hole should be a subtype of 
\rcode{String}. Thus candidates like \textsf{C4} will not be generated as it is
ill-typed (\rcode{arg0} is a \rcode{DataFrame}).

We use filling the remaining hole in \textsf{C5} to illustrate another feature
of \absynthe{}, enumerating finite abstract domains. \absynthe has the upper
bound of this hole at \rcode{DataFrame}, it will substitute all possible
values from \rcode{PyType} that are subtypes of \rcode{DataFrame}. Since there
is only one type \ie \rcode{DataFrame}, it synthesizes expressions of that type
at the hole. For the next candidate \textsf{C6}, again by running abstract
interpreter bounds for \ehole are determined. The \rcode{_} in the \ehole
signifies that \rcode{ColNames} domain still is an abstract variable, while the
types have been concretized. \absynthe can determine bounds for variables only
if the abstract transformers have conditionals (discussed in 
\S~\ref{subsec:formalism:metalang}), not present in \rcode{pd_merge} of 
\rcode{ColNameInterp}.
Running the abstract interpreter eliminates candidate \textsf{C7} as the partial
program will not satisfy the synthesis goal. Eliminating partial programs
removes a family of concrete programs, narrowing the search space further.
\absynthe{} next generates candidates
\textsf{C8} and \textsf{C9}. \textsf{C8}, however, is eliminated
because the \rcode{ColNames} domain interpreter computes the final data frame
will have columns $\{$\rcode{'id'}, \rcode{'valueA'}$\}$. Eventually, the keyword
argument to the merge method is filled with an array. Some ways of filling that
argument fail the test cases (\textsf{C10}), but \textsf{C11} passes all tests
and is accepted as the solution (after being wrapped in a Python method
defintion), also shown in Figure~\ref{fig:synthesized-prog}.

\section{Formalism}
\label{sec:formalism}

In this section we formalize \absynthe in a core language \flang.
Figure~\ref{fig:flang-syn-relations} shows the \flang syntax. Expressions in \flang
have values \val, drawn from a set of \textit{concrete values} \valset;
variables \var; holes $\ehole: \absval$ tagged with an abstraction
\absval; and function application \func{\expr}{\expr}. Note that these are
external functions $f$, e.g., to call out to libraries.
Programs in \flang consist of a single function definition $\eprog{\meth}{\var}
{\expr}$ of a function $\meth$ that takes an argument \var and returns the
result of evaluating \expr.

Abstractions $\absval$ include abstract values $\absconst$ drawn from an
abstract domain $\absvalset$. We assume this domain forms a complete lattice
with greatest element $\top$, least element $\bot$, and partial ordering
$\absval_1 \subseteq \absval_2$. Abstractions also include \emph{abstract
variables}~$\absvar$, which \absynthe{} uses to label holes whose abstractions
cannot immediately be determined. For example, if \absynthe{} synthesizes an
application of a function~$f$, it labels $f$'s arguments with abstract
variables. During synthesis, \absynthe{} maintains bounds on such variables to
narrow down the search space (see below). We refer to abstract values from
\S~\ref{sec:overview} as \emph{abstractions} in this section to avoid the
ambiguity between abstract variables and values.
%
%
Concrete values are lifted to abstract values using the abstraction function
$\alpha$, mapping concrete values to abstract values, \ie $\alpha$ maps \valset
to \absvalset. Likewise, abstract values map to a set of concrete values using
the concretization function $\gamma$, \ie $\gamma$ maps \absvalset{} to the
$\powerset(\valset)$.
We write $\val \in \absconst$ as a shorthand for checking that \val is in the
concretization of $\absconst$. We assume that for each function $f$, we
have a corresponding abstract transfer function $f^\#$ that soundly captures its
semantics.
Finally, during synthesis, \absynthe{} maintains two variable environments:
$\tenv$, binding variables \var to their abstractions, and $\benv$, binding
abstract variables \absvar to their bounds. Abstract variable bounds are written
as a tuple of the lower and upper bound respectively (details in
\S~\ref{subsec:formalism:metalang}).

\begin{figure}
\begin{subfigure}{\textwidth}
\begin{center}
  $$
  \begin{array}{lrcl}
    \emph{Expressions} \quad
    & \expr
    & ::= & \val \spmid \var \spmid \ehole: \absval \spmid
    \func{\expr}{\expr}\\
    \emph{Programs} \quad
    & \program
    & ::= & \eprog{\meth}{\var}{\expr} \\
  \emph{Concrete Values} \quad
  & \val & \in & \valset \\
    \emph{Abstractions} \quad
    & \absval
    & ::= & \absconst  \spmid \absvar \\
  \emph{Abstract Values} \quad
  & \absconst & \in & \absvalset \\ \\

  \emph{Abstraction Function} \quad
  & \alpha & : & \valset \rightarrow \absvalset \\
  \emph{Concretization Function} \quad
  & \gamma & : & \absvalset \rightarrow \powerset(\valset) \\
  \emph{Inclusion} \quad
    & \val \in \absconst
    & \textrm{if} & \val \in \gamma(\absconst)\\
    \emph{Abstract Transfer Function} \quad
  & f^\# & : & (\absvalset, \ldots, \absvalset) \rightarrow \absvalset \\ \\

    \emph{Abstract Environment} \quad
    & \tenv
    & ::= & \emptyset \spmid \var: \absval,~\tenv \\
    \emph{Bounds Environment} \quad
    & \benv
    & ::= & \emptyset \spmid \absvar: (\absval, \absval),~\benv
  \end{array}
  $$
    \caption{Syntax and relations of \flang.}
  \label{fig:flang-syn-relations}
\end{center}
\end{subfigure}




\begin{subfigure}{\textwidth}
  \centering
  \judgementHead{\tenv \vdash \expr \Downarrow \absval}
  \begin{mathpar}
    \inference{\eta(\val) = \absval}
    {\tenv \vdash \val \Downarrow \absval}[\rulename{E-Val}]

    \inference{}{\tenv \vdash \var \Downarrow \tenv[\var]}[\rulename{E-Var}]

  \inference{ }
  {\tenv \vdash \ehole: \absval \Downarrow \absval}[\rulename{E-Hole}]

    \inference{\tenv \vdash \expr_1 \Downarrow \absval_1
    \quad
    \ldots
    \quad
    \tenv \vdash \expr_n \Downarrow \absval_n}
    {\tenv \vdash \func{\expr_1}{\expr_n} \Downarrow \absfunc{\absval_1}{\absval_n}}[\rulename{E-Fun}]
  \end{mathpar}
  \caption{Abstract semantics for \flang.}
  \label{fig:flang-interp-semantics}
\end{subfigure}

  \caption{Syntax, relations, and abstract semantics of \flang.}
  \label{fig:flang-syntax}
\end{figure}

\paragraph{Abstract Semantics}

Next, we define semantics to abstractly interpret candidate programs in our
domain. Figure~\ref{fig:flang-interp-semantics} presents the relation $\tenv
\vdash \expr \Downarrow \absval$ that, given an abstract environment \tenv,
evaluates an expression \expr to an abstract value \absval.
\rulename{E-Val} lifts a concrete value to the abstract
domain by applying the abstraction function. \rulename{E-Var} lifts a variable
to an abstract value by substituting the value from the environment \tenv.
\rulename{E-Hole} abstractly evaluates a hole to its label.
Finally, \rulename{E-Fun} recursively evaluations a function application's
arguments and then applies the abstract transfer function $f^\#$.


%



\paragraph{Synthesis Problem}

We can now formally specify the synthesis problem: Given an abstract
domain~\absvalset, a set of abstract transformers~$f^\#$, and an abstract
specification of the function's input and output $\absval_1 \rightarrow
\absval_2$, synthesize a set of programs~$\program$ such that $\textsc{NoHole}
(\program)$, i.e., $\program$ has no holes in it, and $x:\absval_1, \emptyset
\vdash \program \Downarrow \absval_2$, i.e., $\program$ abstractly evaluates to
$\absval_2$ given that $x$ has abstract value $\absval_1$. Then, the final
solution is chosen as a synthesized candidate
$\program$ that passes all input/output examples.

\subsection{Abstract Transformer Function DSL}
\label{subsec:formalism:metalang}

Figure~\ref{fig:metalang} shows \metalang, the DSL to define abstract
transformer functions $f^\#$ for \absynthe. The primary purpose of the DSL is to
let users define $f^\#$ that can handle both abstract
values \absval and variables \absvar correctly. It is expressive enough to write
the abstract transformer function for domains in \S~\ref{sec:overview}.
Expressions \dslexpr
in \metalang can be either such abstractions \absval, variables \dslvar,
function application \metafun{\dslexpr} and if-then-else statements. We
consider $g$ as uninterpreted abstract functions. The
conditionals \branch for if statements include \textbf{\texttt{top?}} that tests
if an expression is $\top$, \textbf{\texttt{bot?}} that tests
if an expression is $\bot$, \textbf{\texttt{var?}} that tests
if an expression is an abstract variable $\absvar$, and  \textbf{\texttt{val?}}
that tests if an expression is a abstract value $\absval$. Additionally,
expressions \dslexpr can test for ordering using $\subseteq$ or can call an
abstract function \metafun{\dslexpr}. The else branch of these conditionals
evaluate to $\top$, \ie it evaluates to the largest possible abstraction $\top$
if a test of ordering fails. This is done to soundly over-approximate program
behavior, while sacrificing precision.
The abstract transformer is defined as a function $f^\#$ that
takes the input abstract value as argument \dslvar and computes the output
abstraction by evaluating the expression \dslexpr.

\newcommand\foo{\mathit{T}}

\begin{figure}
\begin{subfigure}{\textwidth}
  \centering

  $$
  \begin{array}{lccl}
  \emph{Expressions} \quad
  & \dslexpr
  & ::= & \absval \spmid \dslvar \spmid \metafun{\dslexpr} \spmid 
          \eif{\branch}{\dslexpr}{\dslexpr} \\
  &     & \spmid & \eif{\dslexpr \subseteq \dslexpr}{\dslexpr}{\top} \spmid
          \eif{\metafun{\dslexpr}}{\dslexpr}{\top} \\
  \emph{Conditionals} \quad
  & \branch
  & ::= & \textbf{\texttt{top?}}\ \dslexpr \spmid
          \textbf{\texttt{bot?}}\ \dslexpr \spmid
          \textbf{\texttt{var?}}\ \dslexpr \spmid
          \textbf{\texttt{val?}}\ \dslexpr \\
  \emph{Transfer Functions} \quad
  & \hat{\program}
  & ::= & \eprog{f^\#}{\dslvar}{\dslexpr}
  \end{array}
  $$

  \caption{Syntax of \metalang.}
  \label{fig:metalang}
\end{subfigure}

\begin{subfigure}{\textwidth}
\centering
\judgementHead{\tenv \vdash \langle \benv, \dslexpr \rangle \Downarrow \langle
\benv, \absval \rangle}
\begin{mathpar}
\inference{\tenv[\dslvar] = \absval}{\tenv \vdash \langle \benv, \dslvar \rangle
\Downarrow \langle
\benv, \absval \rangle}[\rulename{A-Var}]

\inference{\tenv \vdash \langle \benv_1, \dslexpr \rangle
\Downarrow \langle \benv_2, \absval \rangle}
{\tenv \vdash \langle \benv_1, \metafun{\dslexpr} \rangle
\Downarrow \langle \benv_2, \metafun{\absval} \rangle} [\rulename{A-Func}]

\inference{\tenv \vdash \langle \benv_1, \branch \rangle \Downarrow
\langle \benv_2, \vtrue \rangle \\
\tenv \vdash \langle \benv_2, \dslexpr_1 \rangle \Downarrow \langle \benv_3, \val
\rangle}
{\tenv \vdash \langle \benv_1, \eif{\branch}{\dslexpr_1}{\dslexpr_2} \rangle
\Downarrow \langle \benv_3, \val \rangle}[\rulename{A-IfT}]


\inference{\tenv \vdash \langle \benv, \dslexpr \rangle
\Downarrow \langle \benv, \top \rangle}
{\tenv \vdash \langle \benv, \textbf{\texttt{top?}} \rangle
\Downarrow \langle \benv, \vtrue \rangle} [\rulename{A-TopT}]



\inference{\tenv \vdash \langle \benv_1, \dslexpr_1 \rangle \Downarrow
\langle \benv_2, \absvar \rangle \\
\tenv \vdash \langle \benv_2, \dslexpr_2 \rangle \Downarrow
\langle \benv_3, \absconst \rangle\\
\benv_3[\absvar] = (\absval_2, \absval_3)
\quad
\absval_2 \subseteq \absconst \subseteq \absval_3\\
\benv_4 = \benv_3[\absvar \mapsto (\absval_2, \absconst)]}
{\tenv \vdash \langle \benv_1, \dslexpr_1 \subseteq \dslexpr_2 \rangle
\Downarrow \langle \benv_4, \vtrue \rangle}
[\rulename{A-VC}]


\inference{\tenv \vdash \langle \benv_1, \dslexpr_1 \rangle \Downarrow
\langle \benv_2, \absvar_1 \rangle \\
\tenv \vdash \langle \benv_2, \dslexpr_2 \rangle \Downarrow
\langle \benv_3, \absvar_2 \rangle}
{\tenv \vdash \langle \benv_1, \dslexpr_1 \subseteq \dslexpr_2 \rangle
\Downarrow \langle \foo(\benv_3, \absvar_1, \absvar_2), \vtrue \rangle}
[\rulename{A-VS}]

\foo(\benv,\absvar_1,\absvar_2)=
\begin{cases}
  \benv[\absvar_1 \mapsto (\absval_3, \absval_4)]
  &\mathit{if }\absval_1 \subseteq \absval_3,  \absval_4 \subseteq \absval_2\\
  \benv[\absvar_1 \mapsto (\absval_3, \absval_2), \absvar_2 \mapsto (\absval_3, \absval_2)]
  &\mathit{if }  \absval_3 \subseteq \absval_2 \subseteq \absval_4 \\
  \benv
  &\mathit{if }\absval_1 \not\subseteq \absval_4\\
  \multicolumn{2}{l}{\mathit{where\ }\benv[\absvar_1] = (\absval_1, \absval_2), \benv[\absvar_2] = (\absval_3, \absval_4)}
\end{cases}

\end{mathpar}
\caption{Selected \metalang evaluation rules.}
\label{fig:metalang-eval}
\end{subfigure}
\caption{Syntax and evaluation rules of \metalang.}
\label{fig:metalang-syn-evals}
\end{figure}


Figure~\ref{fig:metalang-eval} shows selected big-step evaluation rules for the
abstract transformer functions written in \metalang. Under an abstract
environment \tenv and a bounds environment \benv, expression \dslexpr
evaluates
to a new bounds environment and a value \val. In general these rules reflect
standard big step semantics, except for the $\subseteq$ operation, where the
bounds get constrained because of the comparison.
The rule \rulename{A-IfT} evaluates the branch condition \branch and evaluates
$\dslexpr_1$ if it is \vtrue. A similar rule (omitted here) can be written if the
conditional evaluates to \vfalse.
\rulename{A-TopT} checks if the expression \dslexpr evaluates to $\top$. We
omit evaluation rules for the \vfalse case and other branching predicates such
as \textbf{\texttt{bot?}}, \textbf{\texttt{var?}}, and \textbf{\texttt{val?}}
which are similar to \textbf{\texttt{top?}}.



The rules for evaluating $\expr \subseteq \expr$ are most interesting, as these
test for the $\subseteq$ relation while
constraining abstract variables \absvar to the range under which the relation
$\expr \subseteq \expr$ holds.
In general, the abstract variable narrowing reduces the range of
\absvar to a
sound range for that evaluation through $f^\#$. In effect it is finding
satisfiable range for \absvar for that branch.
\rulename{A-VarConst} tests for the
$\subseteq$ relation when $\dslexpr_1$ evaluates to a variable \absvar and
$\dslexpr_2$ evaluates to a values \absconst. In such a case, if \absconst
is within the range of the variable \absvar the term evaluates to \vtrue, while
updating the upper bound of \absvar to \absconst. This narrows the abstract
variables, while still being sound under which the partial order relation
$\subseteq$ holds true.
A similar symmetrical rule exists (omitted here) where the left hand evaluates
to \absconst and right hand evaluates to \absvar.
Finally, \rulename{A-VarSub} gives the rules for comparing two abstract
variables $\absvar_1$ and $\absvar_2$. It uses a metafunction $\foo$ to
describe the cases where $\absvar_1$ is contained in $\absvar_2$, or has some
overlap, or $\absvar_1$ is less than $\absvar_2$.


\subsection{Abstraction-Guided Synthesis}
\label{subsec:formalism:syn-rules}

\begin{figure}
  \centering
  \judgementHead{\benv, \tenv \vdash \expr \rightsquigarrow \expr: \absval}
  \begin{mathpar}
    \inference{
    \val \in \gamma(\absval')
    \quad
    \absval' \subseteq \absval
    }
    {\benv, \tenv \vdash \ehole: \absval \rightsquigarrow \val: \alpha(\val)}[
    \rulename{S-Val}]

    \inference{
    \tenv[\var] = \absval'
    \quad
    \absval' \subseteq \absval
    }{\benv,\tenv \vdash \ehole: \absval \rightsquigarrow \var: \absval'}[
    \rulename{S-Var}]

  \inference{\absfunc{\absvar_1}{\absvar_n} = \absval'
  \quad
  \benv[\absvar_i] = (\absval_{i,1}, \absval_{i,2})\\
  \absval_{i,1} \subseteq \absval_i
  \quad
  \absval_i \subseteq \absval_{i,2}
  \quad
  \absval' \subseteq \absval
  }
  {\benv, \tenv \vdash \ehole: \absval \rightsquigarrow \func{\ehole: \absval_1}
  {\ehole:
  \absval_n}: \absval'}[\rulename{S-Finite}]

  \inference{\benv, \tenv \vdash \ehole: \absvar_1 \rightsquigarrow
  \expr_1: \absvar_1
  \quad
  \ldots
  \quad
  \benv, \tenv \vdash \ehole: \absvar_{n - 1} \rightsquigarrow
  \expr_{n - 1}: \absvar_{n - 1}\\
  \tenv \vdash \expr_1 \Downarrow \absval_1
  \quad
  \ldots
  \quad
  \tenv \vdash \expr_{n - 1} \Downarrow \absval_{n - 1}\\
  \absfunc{\absval_1}{\absval_n} = \absval'
  \quad
  \absval' \subseteq \absval
  }
  {\benv, \tenv \vdash \ehole: \absval \rightsquigarrow \func{\expr_1:
  \absval_1}{\ehole_n:\absval_n}: \absval'}[\rulename{S-Solve}]

  \inference{\absvar_i\ \textrm{and}\ \absvar'\ \textrm{is fresh}}
  {\benv, \tenv \vdash \ehole: \absval \rightsquigarrow \func{\ehole:
  \absvar_1}{\ehole: \absvar_n}: \absvar'}[\rulename{S-Enumer}]
  \end{mathpar}
\caption{Hole replacement rules for \flang.}
\label{fig:flang-synth-semantics}
\end{figure}

To perform abstraction-guided synthesis, \absynthe{} recursively replaces holes
by suitable expressions and then tests fully concretized candidates. Figure~
\ref{fig:flang-synth-semantics} shows the rules for hole replacement. These
rules prove judgments of the form $\benv, \tenv \vdash \expr_1 \rightsquigarrow
\expr_2: \absval$, meaning in bounds environment \benv and abstract environment
$\tenv$, expression $\expr_1$  takes a step by replacing a hole in $\expr_1$ to
yield a new expression $\expr_2$.
In particular, \rulename{S-Val} replaces $\ehole: \absval$ with a value \val
from the concrete set that \absval abstracts. Similarly, \rulename{S-Var}
replaces a hole with a variable that is compatible with the hole's label.

The next few rules are used to generate function applications, or more
generally, any term that may have more holes.
First, \rulename{S-Finite} generates function application when the domain from
which \absval is drawn is finite, \eg a simple type system that without
polymorphic types or first class lambdas, or an effect system as used in
\citet{rbsyn-pldi21}. This rule can produce multiple candidates with each hole
tagged with distinct abstract values from the domain.
Second, for abstract domains with infinite values that can be represented in a
background theory solver, \absynthe applies the \rulename{S-Solve}. If the
function application requires $n$ arguments, only $n - 1$ arguments are
concretized to a term. This gives the
constraint $\absfunc{\absval_1}{\absval_n} = \absval'$ with only one unknown,
$\absval_n$, that can solved for and assigned to the hole. For $f^\#$ to
be lifted to a SMT solver $f^\#$ should also have an interpretation a background
theory supported by the solver.
This is useful for representing predicate abstractions or numeric domains such
as intervals or string lengths (used in \sygus evaluation in
\S~\ref{subsec:eval:sygus-strings}).
Finally, \rulename{S-Enumer}
replaces a hole with a function application with fresh
abstract variables $\absvar_i$ for the arguments and return. Notice there is no
guarantee $f$ will produce a value of the appropriate abstraction. This is
because, while we assume we have an abstract transfer function $f^\#$, we do not
know what abstraction it will compute without concretizing the arguments.
However, unsound partial programs will be eliminated by the abstract interpreter
as discussed below.
Given only forward evaluation semantics and no other information about the
domains, this is best way to construct partial program candidates.
\absynthe can switch between bottom-up synthesis (\rulename{S-Enumer}) and
top-down goal-directed synthesis (rest of the \rulename{S-} rules) depending on
which rule is applied.
While these rules are non-deterministic, the \absynthe
implementation (\S~\ref{sec:implementation}) chooses and applies these rules for
the correct domain in a fixed order to yield solutions.




\paragraph{Synthesis Algorithm}


\begin{algorithm}
\caption{Synthesis of programs that passes a spec $s$}
\label{alg:syn-loop}
\begin{algorithmic}[1]
\Procedure{Generate}{\mthtype{\absval_1}{\absval_2}, maxSize}

  \State \tenv $\gets \lbrack \var \mapsto \absval_1\rbrack$

  \State $\expr_0 \gets \ehole: \absval_2$

  \State workList $\gets \lbrack \expr_0 \rbrack$

  \While{workList is not empty}

    \State $\expr_{curr} \gets \textrm{pop(workList)}$

    \State $\omega_{enumer}$ $\gets \{ \expr_t \mid
    \tenv \vdash \expr_{curr} \rightsquigarrow \expr_t: \absval \}$

    \State $\omega_{valid} \gets \{ \expr_t \in
    \omega_{enumer} \mid \tenv \vdash \expr_t \Downarrow \absval
    \land \absval \subseteq \absval_2 \}$\label{line:abs-elim}

    \State $\omega_{eval} \gets \{ \expr_t \in
    \omega_{valid} \mid \textsc{NoHole}(\expr_t) \}$

    \State $\omega_{rem} \gets \omega_{valid} - \omega_{eval}$

    \ForAll{$\expr_t \in \omega_{eval}$}

      \State \Return{$\expr_t$} \textrm{if} $\textsc{TestProgram}(\expr_t)$
    \EndFor

    \State $\omega_{rem} \gets \{ \expr_t \in
    \omega_{rem} \mid \texttt{size}(\expr_t) \leq \textrm{maxSize}\}$

    \State workList $\gets$ reorder(workList + $\omega_r$)\label{line:reorder}

  \EndWhile

  \State \Return Error: No solution found
\EndProcedure
\end{algorithmic}
\end{algorithm}

Algorithm~\ref{alg:syn-loop} performs abstraction-guided synthesis. The
algorithm uses a work list and combines synthesis
rules for candidate generation with search space pruning based on abstract
interpretation, in addition to testing in a concrete interpreter. The ordering
of programs in the worklist determines the order in which program candidates
are explored (discussed in \S~\ref{sec:implementation}). The synthesis algorithm
starts off with an empty candidate $\expr_0$ as a base expression in the work
list. At every iteration it pops one item from the work list and applies
synthesis rules (Figure~\ref{fig:flang-synth-semantics}) in a non-deterministic
order to produce multiple candidates $\omega_{enumer}$. Each candidate is
abstractly interpreted, and then checked to see if the computed abstraction
satisfies the goal abstraction. If it is satisfied it is added to the set of
valid candidates $\omega_{valid}$
(line~\ref{line:abs-elim}). As partial programs with holes represent a class
of programs, abstractly interpreting these eliminate a class of programs if they
are not included in the goal $\absval_2$.
Thus, the algorithm iterates through partial programs which are 
\emph{sound} with respect to the abstract specification.
Any unsound programs generated by \rulename{S-Enumer} are pruned here.

Finally, all concrete programs $\omega_{eval}$ are
tested in the interpreter to check if a program satisfies all test cases, in
which case it is returned as the solution. The remaining programs $\omega_{rem}$
contain holes, so these can be expanded further by the application of
synthesis rules. Only programs below the maximum size of the search space are
put back into the work list, and the order of the work list is always
based on some domain-specific heuristics (\S~\ref{sec:implementation} discusses
our program ordering).

\section{Implementation}
\label{sec:implementation}

\absynthe is implemented in approximately 3000 lines of Ruby
excluding dependencies.
It is architected as a core library whose interfaces are used to build a
synthesis tool for a problem domain. Additionally, to support solver-backed
domains, we developed a library (\textasciitilde460 lines) to lazily
convert symbolic expressions to Z3 constraints and solve those in an external
process.
\absynthe uses a term enumerator that, at each
step, visits holes in a term and substitutes it with values or subterms
containing more holes applying the rules shown in 
\S~\ref{subsec:formalism:syn-rules}.
\absynthe requires users to define a translation from the ASTs to
the source program and a method that tests a candidate to return if the test
passed or not.
Users may provide a set of constants for the language which are used as
values to be used in the concretization function. In practice, this is useful
when the language has infinite set of terminals (like Python), and selecting
values from the set of constants makes the term generation tractable. For
\autopandas benchmarks, we infer such constants from the data frame row and
column labels (\S~\ref{sec:overview}).

\absynthe explores program candidates in order of their size, preferring smaller
programs first (line~\ref{line:reorder} of
Algorithm~\ref{alg:syn-loop}). We plan to explore other program exploration
order in future work. The synthesis rules
presented in \S~\ref{subsec:formalism:syn-rules} are non-deterministic, however,
our implementation fixes an order of application such rules. It prefers to
synthesize constants and variables followed by function
applications, hashes, arrays, etc. Moreover, based on the definition of abstract
domains (discussed below), it can automatically choose to apply the 
\rulename{S-Finite} or \rulename{S-Solver} rules. If none of these specialized
rules apply, it uses \rulename{S-Enumer} rule to synthesize subterms.

\paragraph{Abstract Domains}

To guide the search, users need to implement an abstract domain. \absynthe
provides a base class---\rcode{AbstractDomain} from which a
programmer can inherit their own abstract domains implementation, like
Figure~\ref{fig:overview-domains-defn}. The base classes come with machinery
that gives built-in
implementation of $\top$, $\bot$, abstract variables \absvar, and supporting
code for partial ordering between these abstract values.
The user has to define how to construct abstract values for that
domain (the \rcode{initialize} method in \S~\ref{sec:overview}), the partial
ordering relation $\subseteq$ between two abstract \emph{values}.
The abstract variable narrowing (\S~\ref{subsec:formalism:metalang}) is
implemented as the $\subseteq$ method in the \rcode{AbstractDomain} base class.
Solver-aided domains (such as string length in
\S~\ref{subsec:eval:sygus-strings}) construct solver terms when
initializing an abstract value, or apply functions that compute abstract values
(including $\cup$ and $\cap$). These terms are checked for satisfiability of
$\absval_1 \subseteq \absval_2$ in the solver when the $\subseteq$ method
is invoked, and any solved abstract variables are assigned to its
holes. If the solver proves the solver term unsatisfiable, the candidate is
eliminated.
The rule \rulename{S-Finite} is applied for domains with finite abstract values
and \rulename{S-Solve} is used for domains whose values can be inferred using an
SMT solver yielding top-down goal-directed synthesis. In case these cannot be
applied, \absynthe falls back to using the \rulename{S-Enumer} rule that is
equivalent to bottom-up term enumeration.
We plan to explore a more ergonomic API for the \absynthe framework in future
work.

\absynthe also provides a \rcode{ProductDomain} class to automatically derive
product domains by combining any user-defined domains as needed. The $\subseteq$
method on \rcode{ProductDomain} returns the conjunction of respective
$\subseteq$ on the individual domains it is composed of.


\paragraph{Abstract Interpreters}

Each abstract domain needs a definition of abstract
semantics, inherited from the \rcode{AbstractInterp} class provided by
\absynthe (as shown in \S~\ref{sec:overview}). All subclasses
override the \rcode{interpret} method that takes as argument the abstract
environment and the AST of the term that is being evaluated. In practice, it is
implemented as evaluating subterms recursively, and then applying the abstract
transformer function written in a subset of Ruby (similar to \metalang in
\S~\ref{subsec:formalism:metalang}) to evaluate the program in the
abstract domain.
A sound interpreter for \rcode{ProductDomain} is derived automatically, by
composing the interpreters of its base domains. More specifically, it evaluates
the term under individual base domains and then combines the results pointwise
into a product.

\paragraph{Concrete Tests}

Any synthesized term without holes that satisfies the abstract specification is
tested by \absynthe in a reference interpreter against concrete test cases.
\absynthe expects the programmer to define a \rcode{test_prog} method
that calls the reference interpreter with the synthesized source program (as a
string in the source language), and returns a boolean to indicate if the tests
passed. The reference interpreter runs the test case, which
in many cases boils down to checking the program against the provided
input/output examples. If the program passes all test cases, it is
considered the correct solution. If the program fails a test, it is discarded.

\paragraph{Optimizations}

In practice, \absynthe uses a min-heap to store a work list of candidates
ordered by their size. This eliminates the reorder step
(Algorithm~\ref{alg:syn-loop} line~\ref{line:reorder}), saving an average cost of
$\mathcal{O}(n\log{}n)$ at each synthesis loop iteration.
Additionally, we found certain common subterms
occur frequently in the same program, \textit{e.g.}, computing the index of the
first space in a string in a \sygus program.
\absynthe caches small terms 
(containing up to one function application)
that do not have any holes to save the cost of synthesizing these small
fragments. Whenever, a hole with compatible abstract value is found, these
fragments are substituted directly without doing the repetitive work of
synthesizing the function application from scratch again (similar to subterm
reuse in \textsc{DryadSynth}~\cite{HuangDryadSynth2020}).
Finally, \absynthe tests a set of predicates against given input/output
examples, to guess a partial program instead of starting from just a \ehole
term. For example, \absynthe has a predicate
that checks if the output is contained in the input, then the output is a
substring of the input. For the \sygus language, if the predicate
\rcode{(str.contains output input)} tests true,
then the partial program is inferred to be
\rcode{(str.substr input $\ehole$ $\ehole$)}. This reduces the
problem complexity by cutting down the search space. Another predicate
\rcode{(str.suffixof output input)} tests if the input ends with output, then it
infers the partial program \rcode{(str.substr input $\ehole$ (str.len input))},
\ie the program is possibly a substring of the input from some index to the end.
We evaluate the performance impact of the latter two optimizations in \S~
\ref{sec:evaluation} (No Template column in Table~\ref{table:sygus-results}).

\paragraph{Limitations}

While \absynthe is a versatile tool to define custom abstract domains and
combine it with testing in a reference interpreter, the approach does have some
limitations.
First, \absynthe only works with forward evaluation rules over
the abstract domain, in contrast to FlashMeta~\cite{flashmeta} that requires 
``inverse semantics'', \ie rules that given a target abstraction computes the
arguments to the abstract transformer. While specifying only the forward
semantics eases the specification burden for users, it require more compute time
to synthesize subterms such as arguments to functions.
Second, while we found product domain useful to combine separate domains, these
domains remain independent through synthesis, unlike predicates where
all defined semantics can be considered at the same time. We plan to explore
methods to make product domains more expressive in future work.
Third, problems where one can define full formal semantics are a better fit for
solver-aided synthesis tools such as Rosette~\cite{TorlakRosette14} or \semgus~
\cite{KimSemGuS21}. We share performance benchmarks on \sygus strings (which
have good solver-aided tools) to give some evidence for
this in our evaluation (\S~\ref{subsec:eval:sygus-strings}).
Notably, solver-aided tools can jointly reason about subterms. In contrast, when
using solver-aided domains, \absynthe concretizes some of the subterms which
requires enumeration through larger number of terms.
Finally, \absynthe falls back on term enumeration when abstract
domains do not provide any more guidance, often leading to combinatorial
explosion for larger terms.

\section{Evaluation}
\label{sec:evaluation}

We evaluate \absynthe by targeting it in variety of domains, to verify it can
synthesize different workloads. The primary motivation is to evaluate the
general applicability of abstract interpretation-guided synthesis to diverse
problems rather than being a state-of-the-art tool at a single synthesis
benchmark suite. The questions we aim to answer in our evaluation are:

\begin{itemize}
  \item How well does \absynthe work for problems traditionally targeted using
  solver-based strategies using the \sygus strings
  benchmark~\cite{alur2017sygus} (\S~\ref{subsec:eval:sygus-strings})? We also
  discuss the performance impact of optimizations and program exploration
  behavior in \absynthe.
  \item Can \absynthe be adapted to an unrelated problem (not handled by
  any tools that solve \sygus benchmarks) where it is difficult
  to write precise formal semantics? We test this by using \absynthe to
  synthesize Python programs that use the Pandas library from the
  \autopandas~\cite{bavishi2019autopandas} benchmark suite
  (\S~\ref{subsec:eval:autopandas}).

\end{itemize}

\subsection{\sygus Strings}
\label{subsec:eval:sygus-strings}

\paragraph{Benchmarks}

To test that \absynthe is a viable approach to synthesize programs that has been
well explored in prior work, we target it on the \sygus strings benchmark
suite~\cite{alur2017sygus}. We believe strings form a good baseline to compare
\absynthe with other synthesis approaches that rely on enumerative search~
\cite{alureusolver2017}, SMT solvers~\cite{reynoldscvc42017}, and abstract
methods directed by solvers~\cite{syngar} (discussed in details in
\S~\ref{sec:related-work}).
In contrast, \absynthe uses only abstract domains with their forward
transformers to guide the search. We \emph{do not expect} \absynthe to
out-perform the past tools, rather to evaluate if it can solve most of the
benchmarks at a lower cost of defining lightweight abstract domains and partial
semantics upfront.

\sygus strings has 22 benchmarks with 4 variants of each---standard (baseline
set of input/output examples), small (fewer examples than standard), long (more
examples than standard), long-repeat (more examples than long with repeated
examples). As our approach is
dependent only on the abstract specification and testing, not on the number of
examples, we show detailed results for the standard version of these
benchmarks. These results generalize to all variants of each benchmark.
As we aim to evaluate how abstraction guided search performs, we exclude any
programs containing branches. Previous work like \rbsyn~\cite{rbsyn-pldi21} and
\textsc{EuSolver}~\cite{alureusolver2017} have used test cases that cover
different paths through a program to do more efficient synthesis of branching
programs. These can be adapted to a system like \absynthe with minor effort.



\absynthe parses the \sygus specification files directly to prepare the
synthesis goal and load the target language. As \sygus does not come with an
official concrete interpreter for programs, we provide one written in Ruby
that is compliant with the \sygus specifications~\cite{RaghothamanU14}.
\absynthe uses this interpreter as a black-box and does not receive any
additional feedback other than the generated \sygus programs satisfied the
input-output examples or not.

\paragraph{Abstract Domains}

We defined the following abstract domains and their semantics to run the
benchmark suite:

\begin{enumerate}
  \item \textbf{String Length.} A solver-aided domain to lift strings to their
  lengths, while lifting integers and booleans without transformation. This
  means the concretization of the abstract value 5 can be the number 5 and the
  set of all strings of length 5, whereas the boolean abstract value \vtrue or
  \vfalse represents identical concrete values.
  \item \textbf{String Prefix.} A domain to represent the set of strings that
  begin with a common prefix. For example, an abstract value with string ``fo''
  is wider than an abstract value with string ``foo'', as the former denotes all
  strings starting with ``fo'' and the latter includes a subset of that, \ie
  strings starting with ``foo''. The $\subseteq$ operation checks if the prefix
  of one string starts with the prefix of the other.
  \item \textbf{String Suffix.} A domain to represent the set of strings that
  end with a common suffix, similar to string prefix domain. The $\subseteq$
  operation checks if the suffix of one string ends with the suffix of the
  other.
\end{enumerate}

These domains were created by looking at the input/output examples in the
synthesis specs, and encoding the simplest partial semantics that guides the
reasoning. For example, a few problems have programs that start with or end with
a string constant. This is how we designed the string prefix and suffix domains
respectively. On the other hand, many problems produce strings of fixed lengths
or the length of the output string is a function of the length of the
input string. The string string length domain expresses semantics constraints
of this kind. As the string length domain is solver-aided, it can handle
symbolic constraints from abstract variables like the string length of a 
substring \rcode{str.substr} operation is \rcode{j - i} where i and j are the
start and end index respectively. Although the string length domain does not
preserve type information, \sygus being a typed language (type-soundness
enforced by the grammar) all programs in the language
are type-correct by construction. Consequently, we did not need to write a type
system as an abstract domain.

Finally, we give abstract specifications in the selected abstract domains where
required. Specifically, we run each benchmark without an abstract annotation,
\ie equivalent to $\top \rightarrow \top$ specification which results in naive
enumeration combined with abstract interpretation. If a benchmark times
out, then we add an abstract annotation, such as $\top \rightarrow 
\textrm{``Dr. ''}$ for the \textsf{dr-name} example
(Table~\ref{table:sygus-results}). This specification means, \absynthe
should find a function that given any input string ($\top$), it computes strings
starting with \textrm{``Dr. ''} only.

\paragraph{Results}

\begin{table}
\centering
\small
\caption{Results of running \absynthe on \sygus strings benchmarks. \textit{\#
Ex} lists the number of I/O examples; \textit{Time} lists the median
and semi-interquartile range for 11 runs; \textit{Size} and \textit{Ht}
reports the number of AST nodes and the height of the program AST respectively; 
\textit{\# Tested} is the number of programs run in the concrete interpreter
before a solution was found; \textit{Domains} lists the domains used to
specify the abstract spec; and \textit{\# Elim} lists the number of partial
programs eliminated by the abstract interpreter during search. \textit{No cache}
and \textit{No Temp} measure the performance of \absynthe
when small expression cache and template inference (\S~\ref{sec:implementation})
are disabled respectively.}
\label{table:sygus-results}
\begin{tabular}{|r|r||r|r|r|r|r|r||r|r|}
\hline
\thead{Benchmark} & \thead{\# Ex} & \thead{Time (sec)}         & \thead{Size} & \thead{Ht} & \thead{\# Tested} & \thead{Domains} & \thead{\# Elim} & \thead{No Cache} & \thead{No Temp} \\
\hline
bikes             & 6             & \btime{1.70}{0.02}   & 7  & 4          & 4808   & $\top$       & 0      & 2.55   & 35.05  \\
dr-name           & 4             & \btime{1.54}{0.02}   & 11 & 4          & 4797   & Prefix       & 46610  & 139.53 & 2.92   \\
firstname         & 4             & \btime{0.03}{0.00}   & 7  & 3          & 4      & $\top$       & 0      & 0.63   & 0.18   \\
initials          & -             & -                    & -  & -          & -      & -            & -      & -      & -      \\
lastname          & 4             & \btime{0.02}{0.00}   & 10 & 4          & 15     & $\top$       & 0      & 0.81   & 18.72  \\
name-combine      & 6             & \btime{0.21}{0.00}   & 5  & 3          & 566    & $\top$       & 0      & 0.24   & 0.22   \\
name-combine-2    & 4             & \btime{6.01}{0.06}   & 9  & 4          & 9723  & Suffix       & 48516  & 6.65   & 8.28   \\
name-combine-3    & 6             & \btime{47.86}{0.23}  & 9  & 5          & 117370 & Suffix       & 124573 & 68.29  & 43.63  \\
name-combine-4    & -             & -                    & -  & -          & -      & -            & -      & -      & -      \\
phone             & 6             & \btime{0.03}{0.00}   & 4  & 2          & 3      & $\top$       & 0      & 0.03   & 0.12   \\
phone-1           & 6             & \btime{0.16}{0.00}   & 6  & 3          & 1189    & $\top$       & 0      & 0.20   & 7.32   \\
phone-2           & 6             & \btime{0.05}{0.01}   & 7  & 3          & 41    & $\top$       & 0      & 0.04   & 63.82  \\
phone-3           & -             & -                    & -  & -          & -      & -            & -      & -      & -      \\
phone-4           & 6             & \btime{0.05}{0.01}   & 4  & 2          & 1577    & $\top$       & 0      & 0.05   & 0.14   \\
phone-5           & 7             & \btime{0.03}{0.00}   & 7  & 3          & 18     & $\top$       & 0      & 2.16   & 0.20   \\
phone-6           & 7             & \btime{100.54}{0.51} & 14 & 4          & 5937   & Length       & 12234  & -      & 27.79  \\
phone-7           & 7             & \btime{103.92}{0.37} & 14 & 4          & 54051   & Length       & 12639  & -      & -      \\
phone-8           & 7             & \btime{0.72}{0.00}   & 10 & 4          & 217     & Length       & 31     & 1.37   & -      \\
phone-9           & -             & -                    & -  & -          & -      & -            & -      & -      & -      \\
phone-10          & -             & -                    & -  & -          & -      & -            & -      & -      & -      \\
reverse-name      & 6             & \btime{0.35}{0.00}   & 5  & 3          & 593    & $\top$       & 0      & 0.41   & 0.42   \\
univ-1            & 6             & \btime{6.69}{0.07}   & 7  & 3          & 19683  & $\top$       & 0      & 8.08   & 7.73   \\
\hline
\end{tabular}
\label{table:sygus-strings}
\end{table}

Table~\ref{table:sygus-results} shows the results of running the \sygus strings
benchmarks through \absynthe with the discussed domains.
The numbers are reported as a median of 11 runs on
a 2016 Macbook Pro with a 2.7GHz Intel Core i7 processor and 16 GB RAM. All
experiments had a timeout of 600 seconds.
In Table~\ref{table:sygus-results},
\textit{Benchmark} column is the name of the problem, \textit{\# Ex} shows the
number of input/output examples. \textit{Time} shows the median running time of
the benchmark along with the semi-interquartile range over 11 runs. The
\textit{Size} and \textit{Ht} columns give the size of the synthesized
program as the count of the AST nodes in the \sygus language and the height of
the synthesized program AST respectively. The \textit{\# Tested} column lists
the number of programs that were tested in the concrete interpreter before a
solution was found. An abstraction that works well reduces this number compared
to a worse abstraction or na\"ive enumeration.
\textit{Domains} column lists the domains used for synthesizing the program.
These domains were provided as a specification in the abstract domain. $\top$
denotes that an abstract specification was provided as a product of $\top$
values in all individual domains for input and output, resulting in just term
enumeration. The rows which mention the domain was provided abstract specs only
from that domain, resulting in guidance from the provided specification.
The \textit{\# Elim} lists the number of partial programs (denotes a family of
concrete programs) that were eliminated by running the abstract interpreter
with the provided specification during the search. For the problems which used
the $\top$ domain, the abstract interpreter did not eliminate any partial
programs, as specification admits all programs.
Any row with \textit{--} denotes time out of the benchmark under these abstract
specifications.

Most benchmarks are solved within $\sim$7 seconds, with
exceptions being \textit{name-combine-3}, \textit{phone-6}, and \textit{phone-7}
which take longer. In general a larger program takes much longer to synthesize,
due to combinatorial increase in the number of terms being searched through as
the AST size increases. For example, larger programs with same AST height take
longer to synthesize due to higher number of function arguments. The number of
examples do not impact the time for
synthesis as most time is spent in abstract interpretation and term generation.
Testing a candidate on the examples take minimal time.
\absynthe performs reasonably well, solving around the same number of benchmarks
as \textsc{EuSolver}~\cite{alur2017sygus}. We selected \textsc{EuSolver} as it is
based on an enumerative search method like \absynthe. The timeout of 600 seconds
only applies to our \absynthe evaluation, whereas \textsc{EuSolver} was
evaluated with a timeout of 3600 seconds. \absynthe solves around
77\% of the benchmarks despite being a tool written a Ruby, one of the more
slower languages. We suspect additional performance gains can be had by writing
the tool in performant language that compiles to native code. We plan to explore
this in future work. Additionally, \absynthe does not have the problem of
overfitting because the search algorithm does not use the input/output examples.
It merely uses it as a test case, and since they do not influence term
enumeration they do not cause overfitting with respect to the examples.

\paragraph{Domain-specific synthesis costs}

Another key advantage of the \absynthe approach is only pay for
what you use. The time of synthesis is dependent on the semantics of the
abstract domain. String prefix and suffix are implemented in pure Ruby and does
not incur much cost for invoking the solver, so these still guide the
search without much cost. However, the string length domain being a
solver-backed domain, requires a call to Z3 for every $\subseteq$ check. So it
give more precise pruning, while taking a longer time for synthesis.
Comparing the average time to generate all
the concrete programs explored gives evidence for this. For example, consider
\textit{phone-6} which explores 5937 candidates in 100.54 seconds (16.93ms
average) with the string length domain, whereas \textit{name-combine-3} explores
117370 candidates in 47.86 seconds (0.41ms average) with the string suffix
domain. Depending on how expensive a domain is, one can combine the domains to
fit in a variety of synthesis time budgets.

\paragraph{Impact of performance optimizations}

We explore the impact of performance optimizations discussed in 
\S~\ref{sec:implementation}. First, the performance of \absynthe on these
benchmarks when the small expressions cache is disabled is reported in the 
\textit{No cache} column. It is slower than the baseline across all benchmarks.
Notably, \textit{phone-6} and \textit{phone-7} reuse function application
subterms. So without caching small expressions, these two benchmarks do
repetitive work synthesizing the same expressions in different call sites,
resulting in a timeout.
Second, the \textit{No Temp} column reports the performance numbers of
\absynthe when it is run on these benchmarks with the template inference by
testing predicates is disabled. It is slower on most benchmarks than the
baseline, and even causing timeouts on some (\textit{phone-7} and 
\textit{phone-8}). The exceptions are \textit{phone-6} and 
\textit{name-combine-3}, where the no templates version is faster than the
baseline. Recall, that the inferred templates have holes, that
have are tagged with a fresh abstract variable \absvar resulting in enumeration
of more terms. In contrast,
the candidate generation rules (\rulename{S-}) applied during the program search
that may potentially synthesize holes with more precise abstractions resulting
in less terms being enumerated. We plan to explore mechanisms to infer template
holes with more precise abstractions in future work.

\subsection{AutoPandas}
\label{subsec:eval:autopandas}

\paragraph{Benchmarks}

We want to test if the approach used by \absynthe, of guiding the search with
lightweight abstract semantics combined with testing to ensure correctness, is
general enough to be useful for another domain. For this purpose we use the
\autopandas~\cite{bavishi2019autopandas} benchmark suite from its artifact
\footnote{GitHub: \url{https://github.com/rbavishi/autopandas}} as a case study.
The benchmarks are sourced from StackOverflow questions containing the 
\texttt{dataframe} tag. Each benchmark contain the input data frames, additional
arguments, the expected data frame output, the list of Pandas API methods to be
used in the program, and the number of method calls in the final program.

~\citet{bavishi2019autopandas} define \emph{smart operators} to generate
candidates and train neural models from a graph-based encoding 
on synthetic data to rank generated candidates.
For a baseline, they consider an enumerative search synthesis engine that
na\"ively enumerates all possible programs using the methods specified in the
benchmark. This narrows down the search space to a permutation of 1, 2, or 3
method calls specified upfront, instead of search over all supported Pandas API.
In contrast, \absynthe works like enumerative search, but large classes of
programs are eliminated by abstract interpretation of partial programs, or
terms are constructed guided by the abstract semantics.
Unlike \sygus, all benchmarks in \autopandas have only one input and output
example. The synthesis goal is a multi-argument Python method that given the
specified input produces the desired output.

The evaluation of \autopandas benchmarks uses the same \absynthe core as the
\sygus evaluation. We wrote a test harness in Python that loads the \autopandas
benchmarks written in Python and communicates with \absynthe core
running as a child process. The \absynthe core is responsible for doing the
enumerative search, while eliminating programs using abstract interpretation.
Any concrete program generated by \absynthe is tested in the host Python
interpreter. These operations are performed as inter-process communication
over Unix pipes between the host Python harness process and the child \absynthe
Ruby process. This allows the testing of generated programs in the host Python
process, saving the overhead of launching a new Python process and importing
Pandas packages (about 1-3 seconds) for every candidate.
If the input/output examples are satisfied the synthesis problem is solved,
else control is returned back to \absynthe which searches and sends the next
candidate for testing.

\paragraph{Abstract Domains}

The abstract domains used for \autopandas benchmarks are:

\begin{enumerate}
  \item \textbf{Types.} A domain to represent the data type of the
  computed values (Figure~\ref{fig:pytype-defn}).
  \item \textbf{Columns.} A domain to represent dataframes as a set of
  their column labels (Figure~\ref{fig:colnames-defn}).
\end{enumerate}

Our Python harness infers the data types and the column labels from the
input/output examples and the \absynthe core constructs the abstract domain
values from \rcode{PyType} and \rcode{ColNames} domains respectively. These
individual domains are combined pointwise using the product domain 
\rcode{PyType $\times$ ColNames}, and \absynthe soundly applies the individual
abstract semantics to compute values in the same product domain.
The types domain in \absynthe is a wrapper around types from
RDL~\cite{rdl-github}, a type system for Ruby. \absynthe uses RDL as a library
to build the \rcode{PyType} class (the \rcode{ty} field holds an RDL type
as shown in Figure~\ref{fig:pytype-defn}). This allows us to reuse prior work
that defines nominal types, generic types, finite hash types, singleton types,
and their subtyping relations. We define the semantics for these RDL types for
the Python language in an abstract interpreter \rcode{PyTypeInterp} to handle
features such as standard method arguments, optional keyword arguments, and
singleton types as arguments (like \rcode{int}).
We define the concretization function $\mu$
over these types, for example, nominal types can be concretized by all
constants of the correct type from the set of constants or the singleton types
are concretized to the singleton value itself.
The semantics of the type domains are defined in terms of
the \rcode{PyType} wrapper that calls into the relevant RDL methods. The example
implementation of these domains in \S~\ref{sec:overview} is a
simplified version of these domains.

In practice, the \autopandas benchmarks have input/output examples that are not
just data frames, but also integers, Python lambdas, and method references
(such as \rcode{nunique} from the Pandas library). \absynthe is soundly able to
abstract these into the relevant domains. For types, integers become 
\rcode{Integer} and lambdas are inferred as a type \rcode{Lambda}. When these
values are lifted to the columns domain, they are represented as $\bot$ as
these are not data frames, thus there is no way to soundly represent their column
labels.
Additionally, \absynthe infers a set of constants from the input/output examples
as well. It adds any string or numeric row and column labels of the data frames,
in addition to any string or numeric standalone values passed as arguments. This
set is used to synthesize the constants during the application of the 
\rulename{S-Val} rules.

\paragraph{Results}

\begin{table}
\centering
\small
\caption{Results of running \autopandas benchmarks through \absynthe. The
\textit{Depth} column shares the longest chain of method calls in the
synthesized solution; \textit{Time} lists the median and semi-interquartile
range of 11 runs for time taken to synthesize a program; \textit{Size} lists
the number of AST nodes in the synthesized solution; \textit{\# Tested}
reports the number of concrete Python programs tested; \textit{AP Neural} and 
\textit{AP Baseline} shares the benchmarks that \autopandas neural model and
na\"ive enumeration could synthesize. The benchmarks denoted with a * were a
part of the artifact, but not reported in the
paper~\cite{bavishi2019autopandas}. Benchmarks highlighted in \hlgreen{blue}
and  \hlred{yellow} shows the benchmarks only synthesized by \absynthe and
\autopandas respectively.}
\label{table:autopandas-results}
\begin{tabular}{|l|r||r|r|r||r|r|}
\hline
\thead{Name} & \thead{Depth} & \thead{Time (sec)} & \thead{Size} & \thead{\# Tested} & \thead{AP Neural} & \thead{AP Baseline} \\
\hline
SO\_11881165              & 1 & \btime{0.20}{0.00}   & 6  & 40      & \checkmark & \checkmark \\
SO\_11941492              & 1 & \btime{13.84}{0.04}  & 5  & 2507    & \checkmark & \checkmark \\
\hlred{SO\_13647222}      & 1 & -                    &    &         & \checkmark & \checkmark \\
\hlgreen{SO\_18172851}    & 1 & \btime{0.42}{0.00}   & 3  & 70      &            &            \\
\hlgreen{SO\_49583055}    & 1 & \btime{3.77}{0.01}   & 6  & 272     &            &            \\
SO\_49592930              & 1 & \btime{0.22}{0.00}   & 3  & 21      & \checkmark & \checkmark \\
SO\_49572546              & 1 & \btime{1.50}{0.01}   & 3  & 548     & \checkmark & \checkmark \\
\hlgreen{SO\_12860421}*   & 1 & \btime{686.50}{1.68} & 11 & 1537521 &            &            \\
SO\_13261175              & 1 & \btime{283.12}{0.39} & 11 & 237755  & \checkmark &            \\
SO\_13793321              & 1 & \btime{5.70}{0.04}   & 6  & 413     & \checkmark & \checkmark \\
SO\_14085517              & 1 & \btime{216.14}{0.38} & 7  & 12844   & \checkmark & \checkmark \\
SO\_11418192              & 2 & \btime{0.10}{0.00}   & 5  & 11      & \checkmark & \checkmark \\
\hlred{SO\_49567723}      & 2 & -                    &    &         & \checkmark &            \\
SO\_49987108*             & 2 & -                    &    &         &            &            \\
SO\_13261691              & 2 & \btime{65.17}{0.17}  & 3  & 22322   & \checkmark & \checkmark \\
SO\_13659881              & 2 & \btime{0.21}{0.00}   & 6  & 45      & \checkmark & \checkmark \\
SO\_13807758              & 2 & \btime{54.92}{0.26}  & 6  & 3144    & \checkmark & \checkmark \\
SO\_34365578              & 2 & -                    &    &         &            &            \\
SO\_10982266              & 3 & -                    &    &         &            &            \\
\hlgreen{SO\_11811392}    & 3 & \btime{6.88}{0.03}   & 4  & 921     &            &            \\
SO\_49581206              & 3 & -                    &    &         &            &            \\
SO\_12065885              & 3 & \btime{0.24}{0.00}   & 6  & 286     & \checkmark & \checkmark \\
\hlred{SO\_13576164}      & 3 & -                    &    &         & \checkmark &            \\
SO\_14023037              & 3 & -                    &    &         &            &            \\
SO\_53762029              & 3 & \btime{545.62}{0.91} & 9  & 229233  & \checkmark & \checkmark \\
\hlred{SO\_21982987}      & 3 & -                    &    &         & \checkmark & \checkmark \\
SO\_39656670              & 3 & -                    &    &         &            &            \\
SO\_23321300              & 3 & -                    &    &         &            &            \\
\hline
\end{tabular}
\end{table}

Table~\ref{table:autopandas-results} shows the results of running the
\autopandas benchmarks through \absynthe. The numbers are collected on a 2016
Macbook Pro with a 2.7GHz Intel Core i7 processor and 16 GB
RAM, with a timeout of 20 minutes (consistent with the timeout
of~\citet{bavishi2019autopandas}). The 
\emph{Name} column shows the name of the benchmark, \ie the
StackOverflow question ID from which the problem is taken. The \emph{Depth}
column shows the length of sequence of method call chain in the final
solution. The \autopandas benchmarks are tuned to synthesize programs with a
chain of method calls, where the bulk of the time spend is in synthesizing
arguments to these method calls. This is characteristic of the Pandas API which
accepts many arguments, often optional keyword arguments.
The \emph{Time} column shows the median of 11 runs along with the
semi-interquartile range, where -- denotes that a benchmark timed out.
The \emph{Size} lists the synthesized program size as number of AST nodes. Note
that, this number is affected by both the depth of the synthesized program 
(the number of method calls) and the number of arguments to those methods.
\emph{\# Tested} lists the number of concrete programs generated by \absynthe
that were tested in the Python interpreter. Finally, \emph{AP Neural} and 
\emph{AP Baseline} shares the benchmarks solved by the \autopandas neural
model and na\"ive enumeration to aid in comparison with \absynthe. Two
benchmarks, \emph{SO\_12860421} and \emph{SO\_49567723}, are marked with a * as
these were found in the \autopandas artifact were not reported in the paper.

\absynthe solves 17 programs, the same number of programs as
\autopandas neural model. However, the set of synthesized programs by both tools
are different with a significant overlap. Benchmarks listed in 
Table~\ref{table:autopandas-results} without any highlight shows the benchmarks
that were synthesized by both tools. Benchmarks highlighted in \hlgreen{blue}
were synthesized only by \absynthe but not by \autopandas. Likewise, benchmarks
highlighted in \hlred{yellow} are the benchmarks synthesized only by \autopandas
but not by \absynthe. 
The time taken to synthesize the programs is largely dictated by how the
abstract semantics prunes the space of programs, hence it is proportional
to the number of concrete programs generated and tested.
The fact that, for the same program size, the number of AST nodes in the method
arguments (the difference between size and depth) is indicative of solving
time shows that synthesizing arguments is indeed the bottleneck of this
benchmark suite. For example, like \emph{SO\_11811392} and \emph{SO\_12065885}
the type system quickly narrows down the search space, and the solution uses API
methods that have 0 or 1 arguments only, making the arguments synthesis quick.


\paragraph{Discussion}

\absynthe solves a harder synthesis problem because it does not use the list of
methods to be used as provided in the specification. Instead, \absynthe uses the
complete set of 30 supported Pandas API for every benchmark. Approximately, this
gives us a choice of permutations of size 1, 2, or 3 (depending on the depth of
the final solution) from 30 methods, without considering arguments from those
methods. In contrast, the baseline enumerative search \emph{AP Baseline}
comparison limits the search to only the Pandas API methods that will be used in
the final solution. Typically this limits the search space to 1, 2, or 3 methods
as given in the specification. In other words, under na\"ive enumeration,
\absynthe explores a strictly larger set of programs than \autopandas baseline.

In the benchmarks where \absynthe failed to synthesize a solution, it falls back
to term enumeration as the abstract domain was not precise. More
specifically in the benchmarks with depth 3, \absynthe could do better by
jointly reasoning about values in relational abstractions between multiple
arguments of the same method. We plan to explore support for relational
abstractions in future work. The neural model trained by 
\citet{bavishi2019autopandas} is good at guessing the
sequences that are potentially likely to solve the synthesis task. It, however,
does not take into account semantics of the program, thus eliminating impossible
programs from being considered. This shows up in \emph{SO\_18172851} and
\emph{SO\_49583055} where both enumerative search and neural models failed, but
\absynthe succeeds. Moreover, any updates to the neural model would need to be
addressed with a new encoding or a retraining of the model on new data, a
potentially resource consuming process. However, exploring the synergy of
guidance from abstract interpretation combined with neural models similar to
\citet{AndersonVDC20} to rank \emph{sound} program candidate choices is an
interesting future work.



\section{Related Work}
\label{sec:related-work}

\paragraph{General Purpose Synthesis Tools}

\semgus~\cite{KimSemGuS21} has the same motivation as \absynthe to
develop a general-purpose abstraction guided synthesis framework. However,
\semgus requires the programmer to provide semantics in a relational format as
constrained horn clauses (CHCs). While CHCs are expressive and have dedicated
solvers~\cite{KomuravelliSpacer2016}, correctly defining semantics as a
relations is prohibitively time-consuming and error-prone. Moreover, \semgus
performs well in proving unrealizability of synthesis problems, but it has
limited success in synthesizing solutions. In contrast, \absynthe is a
dedicated synthesizer that is geared towards synthesizing programs based on
executable abstract semantics. \absynthe can be thought of as an unrealizability
prover if coarse-grained semantics, the focus of \absynthe, is sufficient to
prove unrealizable. \semgus also supports under-approximate semantics, which is
an interesting future work in the context of \absynthe.
Rosette~\cite{TorlakRosette14} and Sketch~\cite{SolarLezamaSketch13} are
solver-aided languages that use bounded verification using a SMT solver to
synthesize programs written in a DSL. In contrast, \absynthe relies on abstract
interpretation to guide search, so it can reason about unbounded
program properties.
There has been parallel work in synthesis using Christiansen
grammars~\cite{OrtegaCA07} that allows one to encode some program semantics as
context-dependent properties directly in the syntax grammar. However, an
abstract interpreter-based approach gives \absynthe more semantic reasoning
capabilities (like polymorphism).

\paragraph{Domain-specific synthesis}

\sygus~\cite{AlurBJMRSSSTU13} being a standard synthesis problem specification
format, has seen a variety of solver approaches. CVC4~\cite{reynoldscvc42017} is
a general-purpose SMT solver that has support for synthesizing programs in the
\sygus format.
CVC4 has complete support for theory of strings and linear integer arithmetic,
so it performs better than \absynthe (which is guided by simple abstract
domains) for \sygus. However, \absynthe's strength is generalizability to other
kinds of synthesis problems as demonstrated in synthesis of \autopandas
benchmarks (\S~\ref{subsec:eval:autopandas}).
DryadSynth~\cite{HuangDryadSynth2020} explores a reconciling
deductive and enumerative synthesis in \sygus problems limited to the
conditional linear integer arithmetic background theory. Some of their findings
has been adopted by \absynthe (\S~\ref{sec:implementation}).
\textsc{EuSolver}~\cite{alureusolver2017} is an enumerative solver that takes a
divide-and-conquer approach. It synthesizes individual programs that are correct
on a subset of examples, and predicates that distinguishes the program and
combines these into a single final solution. \absynthe is close to 
\textsc{EuSolver}, as it is also based on enumerative search, but it is also
guided by abstract semantics as well. We plan to support synthesizing
conditionals in future work.

Past work solves synthesis problems using domain specific abstractions such as
types and examples~\cite{osera2015myth,frankle2016example}, over-approximate
semantics on table operations~\cite{fengtable2017}, refinement
types~\cite{Polikarpova2016Synquid}, secure
declassification~\cite{guria2022anosy}, abstract domain to verify atomic sections of a program~\cite{VechevYY10},
and SQL equivalence relations~\cite{Wang2017Scythe}. These abstraction can be
designed as a domain and an abstract evaluation semantics can be provided to
\absynthe for synthesizing such programs. However, \absynthe being a general
purpose synthesis tool, will not have domain specific optimizations. We plan to
explore \absynthe as platform deploying domain specific synthesis in
future work.

\paragraph{Abstraction-guided Synthesis}

\textsc{Simpl}~\cite{sosimpl2017} combines enumerative search with static
analysis based pruning, which is similar to \absynthe. However, the program
search in \absynthe can be parameterized by a user provided abstract
interpreter allowing the user to write specifications and semantics in a domain
fit for the task-at-hand. Additionally, \absynthe can infer abstract values for
the holes in partial programs, thus guiding the search using the abstract
semantics (Figure~\ref{fig:flang-synth-semantics}).
\textsc{Blaze}~\cite{syngar} is very similar to \absynthe as it uses abstract
semantics to guide the search. It adapts \emph{counterexample guided
abstraction refinement} to synthesis problems by refining the abstraction when a
test fails, and constructing a proof of incorrectness in the process. However,
it starts with a universe of predicates that is used for abstraction
refinement, which is a requirement \absynthe doesn't place on users.
FlashMeta~\cite{flashmeta} is similar, but requires the definition of
``inverse'' semantics for operators using \emph{witness functions}. \absynthe,
however, requires only the definition of forward abstract semantics
and attempts to derive the inverse semantics automatically where possible.

\paragraph{Learning-based approaches}

There has been a recent rise of learning based approaches to make
program synthesis more tractable. \autopandas~\cite{bavishi2019autopandas} is an
example of applying neural models to rank candidate choices constructed by other
program generation methods (\emph{smart operators} in \autopandas' case).
\textsc{DeepCoder}~\cite{BalogGBNT17} trains a deep neural network to predict
properties of programs based on input/output examples. These properties are used
to augment the search by an enumerative search or SMT solvers. \absynthe is
complementary to these approaches and does not use machine learning. In
future, we plan to explore extensions to \absynthe that reorders the
program search order using a model learned on program text \emph{and}
abstract semantics.
\textsc{EuPhony}~\cite{lee-euphony-2018}, on the other hand, uses an approach
inspired by transfer learning to learn a \emph{probabilistic higher order
grammar}, and uses
that in enumerative search to synthesize solutions. \textsc{Probe}~
\cite{BarkeProbe20} learn a probabilistic grammar \emph{just-in-time} during
synthesis. Their key insight is
that many \sygus programs that pass a few examples have parts of the syntax that
has higher likelihood to be present in the final solution. In contrast,
\absynthe is complementary to the approach of learning probabilistic grammars;
abstract domains can prune the space of programs, while the grammar can assign
higher weights to the terms that should be enumerated earlier. We leave
exploring the synergy between these approaches to future work.


\section{Conclusion}

We presented \absynthe, a tool that combines abstract
interpretation and testing to synthesize programs. It accepts user-defined
lightweight abstract domains and partial semantics for the language as an input,
and enables guided search over the space of programs in the language. We
evaluated \absynthe on \sygus strings benchmarks and found \absynthe can solve
77\% of the benchmarks, most within 7 seconds. Moreover, \absynthe supports a
pay-as-you-go model, where the user only pays for the abstract domain they are
using for synthesis. Finally, to evaluate the generality of \absynthe to other
domains, we use it to synthesize Pandas data frame manipulation programs in
Python from the \autopandas benchmark suite. \absynthe performs at par with
\autopandas and synthesizes programs with low specification burden, but no
neural network training costs. We believe \absynthe demonstrates a promising
design choice for design of synthesis tools that leverage testing for
correctness along with lightweight abstractions with partial semantics for
search guidance.

\section*{Data Availability Statement}

The latest version of the tool \absynthe is publicly available on GitHub
\footnote{\url{https://github.com/ngsankha/absynthe}}. A snapshot of \absynthe,
along with source code, benchmarks used in the paper, supporting scripts and
instructions to reproduce our results in \S~\ref{sec:evaluation} are available
as a Docker image artifact~\cite{artifact}.

\begin{acks}
Thanks to the anonymous reviewers for their helpful comments. This research
was supported in part by National Science Foundation awards \#1900563 and
\#1846350.
\end{acks}

\bibliography{references}


\begin{thebibliography}{37}


\ifx \showCODEN    \undefined \def \showCODEN     #1{\unskip}     \fi
\ifx \showDOI      \undefined \def \showDOI       #1{#1}\fi
\ifx \showISBNx    \undefined \def \showISBNx     #1{\unskip}     \fi
\ifx \showISBNxiii \undefined \def \showISBNxiii  #1{\unskip}     \fi
\ifx \showISSN     \undefined \def \showISSN      #1{\unskip}     \fi
\ifx \showLCCN     \undefined \def \showLCCN      #1{\unskip}     \fi
\ifx \shownote     \undefined \def \shownote      #1{#1}          \fi
\ifx \showarticletitle \undefined \def \showarticletitle #1{#1}   \fi
\ifx \showURL      \undefined \def \showURL       {\relax}        \fi
\providecommand\bibfield[2]{#2}
\providecommand\bibinfo[2]{#2}
\providecommand\natexlab[1]{#1}
\providecommand\showeprint[2][]{arXiv:#2}

\bibitem[Alur et~al\mbox{.}(2013)]%
        {AlurBJMRSSSTU13}
\bibfield{author}{\bibinfo{person}{Rajeev Alur}, \bibinfo{person}{Rastislav
  Bod{\'{\i}}k}, \bibinfo{person}{Garvit Juniwal}, \bibinfo{person}{Milo M.~K.
  Martin}, \bibinfo{person}{Mukund Raghothaman}, \bibinfo{person}{Sanjit~A.
  Seshia}, \bibinfo{person}{Rishabh Singh}, \bibinfo{person}{Armando
  Solar{-}Lezama}, \bibinfo{person}{Emina Torlak}, {and}
  \bibinfo{person}{Abhishek Udupa}.} \bibinfo{year}{2013}\natexlab{}.
\newblock \showarticletitle{Syntax-guided synthesis}. In
  \bibinfo{booktitle}{\emph{Formal Methods in Computer-Aided Design, {FMCAD}
  2013, Portland, OR, USA, October 20-23, 2013}}. \bibinfo{publisher}{{IEEE}},
  \bibinfo{pages}{1--8}.
\newblock
\urldef\tempurl%
\url{https://ieeexplore.ieee.org/document/6679385/}
\showURL{%
\tempurl}


\bibitem[Alur et~al\mbox{.}(2017a)]%
        {alur2017sygus}
\bibfield{author}{\bibinfo{person}{Rajeev Alur}, \bibinfo{person}{Dana Fisman},
  \bibinfo{person}{Rishabh Singh}, {and} \bibinfo{person}{Armando
  Solar{-}Lezama}.} \bibinfo{year}{2017}\natexlab{a}.
\newblock \showarticletitle{SyGuS-Comp 2017: Results and Analysis}. In
  \bibinfo{booktitle}{\emph{Proceedings Sixth Workshop on Synthesis, SYNT@CAV
  2017, Heidelberg, Germany, 22nd July 2017}} \emph{(\bibinfo{series}{{EPTCS}},
  Vol.~\bibinfo{volume}{260})}. \bibinfo{pages}{97--115}.
\newblock
\urldef\tempurl%
\url{https://doi.org/10.4204/EPTCS.260.9}
\showDOI{\tempurl}


\bibitem[Alur et~al\mbox{.}(2017b)]%
        {alureusolver2017}
\bibfield{author}{\bibinfo{person}{Rajeev Alur}, \bibinfo{person}{Arjun
  Radhakrishna}, {and} \bibinfo{person}{Abhishek Udupa}.}
  \bibinfo{year}{2017}\natexlab{b}.
\newblock \showarticletitle{Scaling Enumerative Program Synthesis via Divide
  and Conquer}. In \bibinfo{booktitle}{\emph{Tools and Algorithms for the
  Construction and Analysis of Systems - 23rd International Conference, {TACAS}
  2017, Held as Part of the European Joint Conferences on Theory and Practice
  of Software, {ETAPS} 2017, Uppsala, Sweden, April 22-29, 2017, Proceedings,
  Part {I}}} \emph{(\bibinfo{series}{Lecture Notes in Computer Science},
  Vol.~\bibinfo{volume}{10205})}, \bibfield{editor}{\bibinfo{person}{Axel
  Legay} {and} \bibinfo{person}{Tiziana Margaria}} (Eds.).
  \bibinfo{pages}{319--336}.
\newblock
\urldef\tempurl%
\url{https://doi.org/10.1007/978-3-662-54577-5\_18}
\showDOI{\tempurl}


\bibitem[Anderson et~al\mbox{.}(2020)]%
        {AndersonVDC20}
\bibfield{author}{\bibinfo{person}{Greg Anderson}, \bibinfo{person}{Abhinav
  Verma}, \bibinfo{person}{Isil Dillig}, {and} \bibinfo{person}{Swarat
  Chaudhuri}.} \bibinfo{year}{2020}\natexlab{}.
\newblock \showarticletitle{Neurosymbolic Reinforcement Learning with Formally
  Verified Exploration}. In \bibinfo{booktitle}{\emph{Advances in Neural
  Information Processing Systems 33: Annual Conference on Neural Information
  Processing Systems 2020, NeurIPS 2020, December 6-12, 2020, virtual}},
  \bibfield{editor}{\bibinfo{person}{Hugo Larochelle},
  \bibinfo{person}{Marc'Aurelio Ranzato}, \bibinfo{person}{Raia Hadsell},
  \bibinfo{person}{Maria{-}Florina Balcan}, {and} \bibinfo{person}{Hsuan{-}Tien
  Lin}} (Eds.).
\newblock
\urldef\tempurl%
\url{https://proceedings.neurips.cc/paper/2020/hash/448d5eda79895153938a8431919f4c9f-Abstract.html}
\showURL{%
\tempurl}


\bibitem[Balog et~al\mbox{.}(2017)]%
        {BalogGBNT17}
\bibfield{author}{\bibinfo{person}{Matej Balog}, \bibinfo{person}{Alexander~L.
  Gaunt}, \bibinfo{person}{Marc Brockschmidt}, \bibinfo{person}{Sebastian
  Nowozin}, {and} \bibinfo{person}{Daniel Tarlow}.}
  \bibinfo{year}{2017}\natexlab{}.
\newblock \showarticletitle{DeepCoder: Learning to Write Programs}. In
  \bibinfo{booktitle}{\emph{5th International Conference on Learning
  Representations, {ICLR} 2017, Toulon, France, April 24-26, 2017, Conference
  Track Proceedings}}. \bibinfo{publisher}{OpenReview.net}.
\newblock
\urldef\tempurl%
\url{https://openreview.net/forum?id=ByldLrqlx}
\showURL{%
\tempurl}


\bibitem[Barke et~al\mbox{.}(2020)]%
        {BarkeProbe20}
\bibfield{author}{\bibinfo{person}{Shraddha Barke}, \bibinfo{person}{Hila
  Peleg}, {and} \bibinfo{person}{Nadia Polikarpova}.}
  \bibinfo{year}{2020}\natexlab{}.
\newblock \showarticletitle{Just-in-time learning for bottom-up enumerative
  synthesis}.
\newblock \bibinfo{journal}{\emph{Proc. {ACM} Program. Lang.}}
  \bibinfo{volume}{4}, \bibinfo{number}{{OOPSLA}} (\bibinfo{year}{2020}),
  \bibinfo{pages}{227:1--227:29}.
\newblock
\urldef\tempurl%
\url{https://doi.org/10.1145/3428295}
\showDOI{\tempurl}


\bibitem[Bavishi et~al\mbox{.}(2019)]%
        {bavishi2019autopandas}
\bibfield{author}{\bibinfo{person}{Rohan Bavishi}, \bibinfo{person}{Caroline
  Lemieux}, \bibinfo{person}{Roy Fox}, \bibinfo{person}{Koushik Sen}, {and}
  \bibinfo{person}{Ion Stoica}.} \bibinfo{year}{2019}\natexlab{}.
\newblock \showarticletitle{AutoPandas: neural-backed generators for program
  synthesis}.
\newblock \bibinfo{journal}{\emph{Proc. {ACM} Program. Lang.}}
  \bibinfo{volume}{3}, \bibinfo{number}{{OOPSLA}} (\bibinfo{year}{2019}),
  \bibinfo{pages}{168:1--168:27}.
\newblock
\urldef\tempurl%
\url{https://doi.org/10.1145/3360594}
\showDOI{\tempurl}


\bibitem[Cousot and Cousot(1977)]%
        {CousotC77}
\bibfield{author}{\bibinfo{person}{Patrick Cousot} {and}
  \bibinfo{person}{Radhia Cousot}.} \bibinfo{year}{1977}\natexlab{}.
\newblock \showarticletitle{Abstract Interpretation: {A} Unified Lattice Model
  for Static Analysis of Programs by Construction or Approximation of
  Fixpoints}. In \bibinfo{booktitle}{\emph{Conference Record of the Fourth
  {ACM} Symposium on Principles of Programming Languages, Los Angeles,
  California, USA, January 1977}}. \bibinfo{publisher}{{ACM}},
  \bibinfo{pages}{238--252}.
\newblock
\urldef\tempurl%
\url{https://doi.org/10.1145/512950.512973}
\showDOI{\tempurl}


\bibitem[Cousot et~al\mbox{.}(2005)]%
        {CousotCFMMMR05}
\bibfield{author}{\bibinfo{person}{Patrick Cousot}, \bibinfo{person}{Radhia
  Cousot}, \bibinfo{person}{J{\'{e}}r{\^{o}}me Feret}, \bibinfo{person}{Laurent
  Mauborgne}, \bibinfo{person}{Antoine Min{\'{e}}}, \bibinfo{person}{David
  Monniaux}, {and} \bibinfo{person}{Xavier Rival}.}
  \bibinfo{year}{2005}\natexlab{}.
\newblock \showarticletitle{The ASTRE{\'{E}} Analyzer}. In
  \bibinfo{booktitle}{\emph{Programming Languages and Systems, 14th European
  Symposium on Programming,ESOP 2005, Held as Part of the Joint European
  Conferences on Theory and Practice of Software, {ETAPS} 2005, Edinburgh, UK,
  April 4-8, 2005, Proceedings}} \emph{(\bibinfo{series}{Lecture Notes in
  Computer Science}, Vol.~\bibinfo{volume}{3444})}.
  \bibinfo{publisher}{Springer}, \bibinfo{pages}{21--30}.
\newblock
\urldef\tempurl%
\url{https://doi.org/10.1007/978-3-540-31987-0\_3}
\showDOI{\tempurl}


\bibitem[Feng et~al\mbox{.}(2018)]%
        {FengMBD18}
\bibfield{author}{\bibinfo{person}{Yu Feng}, \bibinfo{person}{Ruben Martins},
  \bibinfo{person}{Osbert Bastani}, {and} \bibinfo{person}{Isil Dillig}.}
  \bibinfo{year}{2018}\natexlab{}.
\newblock \showarticletitle{Program synthesis using conflict-driven learning}.
  In \bibinfo{booktitle}{\emph{Proceedings of the 39th {ACM} {SIGPLAN}
  Conference on Programming Language Design and Implementation, {PLDI} 2018,
  Philadelphia, PA, USA, June 18-22, 2018}}. \bibinfo{publisher}{{ACM}},
  \bibinfo{pages}{420--435}.
\newblock
\urldef\tempurl%
\url{https://doi.org/10.1145/3192366.3192382}
\showDOI{\tempurl}


\bibitem[Feng et~al\mbox{.}(2017)]%
        {fengtable2017}
\bibfield{author}{\bibinfo{person}{Yu Feng}, \bibinfo{person}{Ruben Martins},
  \bibinfo{person}{Jacob~Van Geffen}, \bibinfo{person}{Isil Dillig}, {and}
  \bibinfo{person}{Swarat Chaudhuri}.} \bibinfo{year}{2017}\natexlab{}.
\newblock \showarticletitle{Component-based synthesis of table consolidation
  and transformation tasks from examples}. In
  \bibinfo{booktitle}{\emph{Proceedings of the 38th {ACM} {SIGPLAN} Conference
  on Programming Language Design and Implementation, {PLDI} 2017, Barcelona,
  Spain, June 18-23, 2017}}. \bibinfo{publisher}{{ACM}},
  \bibinfo{pages}{422--436}.
\newblock
\urldef\tempurl%
\url{https://doi.org/10.1145/3062341.3062351}
\showDOI{\tempurl}


\bibitem[Foster et~al\mbox{.}(2020)]%
        {rdl-github}
\bibfield{author}{\bibinfo{person}{Jeffrey Foster}, \bibinfo{person}{Brianna
  Ren}, \bibinfo{person}{Stephen Strickland}, \bibinfo{person}{Alexander Yu},
  \bibinfo{person}{Milod Kazerounian}, {and} \bibinfo{person}{Sankha~Narayan
  Guria}.} \bibinfo{year}{2020}\natexlab{}.
\newblock \bibinfo{booktitle}{\emph{{RDL: Types, type checking, and contracts
  for Ruby}}}.
\newblock
\urldef\tempurl%
\url{https://github.com/tupl-tufts/rdl}
\showURL{%
\tempurl}


\bibitem[Frankle et~al\mbox{.}(2016)]%
        {frankle2016example}
\bibfield{author}{\bibinfo{person}{Jonathan Frankle},
  \bibinfo{person}{Peter{-}Michael Osera}, \bibinfo{person}{David Walker},
  {and} \bibinfo{person}{Steve Zdancewic}.} \bibinfo{year}{2016}\natexlab{}.
\newblock \showarticletitle{Example-directed synthesis: a type-theoretic
  interpretation}. In \bibinfo{booktitle}{\emph{Proceedings of the 43rd Annual
  {ACM} {SIGPLAN-SIGACT} Symposium on Principles of Programming Languages,
  {POPL} 2016, St. Petersburg, FL, USA, January 20 - 22, 2016}}.
  \bibinfo{publisher}{{ACM}}, \bibinfo{pages}{802--815}.
\newblock
\urldef\tempurl%
\url{https://doi.org/10.1145/2837614.2837629}
\showDOI{\tempurl}


\bibitem[Gulwani(2011)]%
        {Gulwani11}
\bibfield{author}{\bibinfo{person}{Sumit Gulwani}.}
  \bibinfo{year}{2011}\natexlab{}.
\newblock \showarticletitle{Automating string processing in spreadsheets using
  input-output examples}. In \bibinfo{booktitle}{\emph{Proceedings of the 38th
  {ACM} {SIGPLAN-SIGACT} Symposium on Principles of Programming Languages,
  {POPL} 2011, Austin, TX, USA, January 26-28, 2011}},
  \bibfield{editor}{\bibinfo{person}{Thomas Ball} {and} \bibinfo{person}{Mooly
  Sagiv}} (Eds.). \bibinfo{publisher}{{ACM}}, \bibinfo{pages}{317--330}.
\newblock
\urldef\tempurl%
\url{https://doi.org/10.1145/1926385.1926423}
\showDOI{\tempurl}


\bibitem[Guria et~al\mbox{.}(2023)]%
        {artifact}
\bibfield{author}{\bibinfo{person}{Sankha~Narayan Guria},
  \bibinfo{person}{Jeffrey~S. Foster}, {and} \bibinfo{person}{David~Van Horn}.}
  \bibinfo{year}{2023}\natexlab{}.
\newblock \bibinfo{booktitle}{\emph{{Artifact for "Absynthe: Abstract
  Interpretation- Guided Synthesis"}}}.
\newblock
\urldef\tempurl%
\url{https://doi.org/10.5281/zenodo.7824175}
\showDOI{\tempurl}


\bibitem[Guria et~al\mbox{.}(2021)]%
        {rbsyn-pldi21}
\bibfield{author}{\bibinfo{person}{Sankha~Narayan Guria},
  \bibinfo{person}{Jeffrey~S. Foster}, {and} \bibinfo{person}{David Van{
  }Horn}.} \bibinfo{year}{2021}\natexlab{}.
\newblock \showarticletitle{RbSyn: type- and effect-guided program synthesis}.
  In \bibinfo{booktitle}{\emph{{PLDI} '21: 42nd {ACM} {SIGPLAN} International
  Conference on Programming Language Design and Implementation, Virtual Event,
  Canada, June 20-25, 2021}}. \bibinfo{publisher}{{ACM}},
  \bibinfo{pages}{344--358}.
\newblock
\urldef\tempurl%
\url{https://doi.org/10.1145/3453483.3454048}
\showDOI{\tempurl}


\bibitem[Guria et~al\mbox{.}(2022)]%
        {guria2022anosy}
\bibfield{author}{\bibinfo{person}{Sankha~Narayan Guria}, \bibinfo{person}{Niki
  Vazou}, \bibinfo{person}{Marco Guarnieri}, {and} \bibinfo{person}{James
  Parker}.} \bibinfo{year}{2022}\natexlab{}.
\newblock \showarticletitle{{ANOSY:} approximated knowledge synthesis with
  refinement types for declassification}. In \bibinfo{booktitle}{\emph{{PLDI}
  '22: 43rd {ACM} {SIGPLAN} International Conference on Programming Language
  Design and Implementation, San Diego, CA, USA, June 13 - 17, 2022}},
  \bibfield{editor}{\bibinfo{person}{Ranjit Jhala} {and} \bibinfo{person}{Isil
  Dillig}} (Eds.). \bibinfo{publisher}{{ACM}}, \bibinfo{pages}{15--30}.
\newblock
\urldef\tempurl%
\url{https://doi.org/10.1145/3519939.3523725}
\showDOI{\tempurl}


\bibitem[Huang et~al\mbox{.}(2020)]%
        {HuangDryadSynth2020}
\bibfield{author}{\bibinfo{person}{Kangjing Huang}, \bibinfo{person}{Xiaokang
  Qiu}, \bibinfo{person}{Peiyuan Shen}, {and} \bibinfo{person}{Yanjun Wang}.}
  \bibinfo{year}{2020}\natexlab{}.
\newblock \showarticletitle{Reconciling enumerative and deductive program
  synthesis}. In \bibinfo{booktitle}{\emph{Proceedings of the 41st {ACM}
  {SIGPLAN} International Conference on Programming Language Design and
  Implementation, {PLDI} 2020, London, UK, June 15-20, 2020}},
  \bibfield{editor}{\bibinfo{person}{Alastair~F. Donaldson} {and}
  \bibinfo{person}{Emina Torlak}} (Eds.). \bibinfo{publisher}{{ACM}},
  \bibinfo{pages}{1159--1174}.
\newblock
\urldef\tempurl%
\url{https://doi.org/10.1145/3385412.3386027}
\showDOI{\tempurl}


\bibitem[Kim et~al\mbox{.}(2021)]%
        {KimSemGuS21}
\bibfield{author}{\bibinfo{person}{Jinwoo Kim}, \bibinfo{person}{Qinheping Hu},
  \bibinfo{person}{Loris D'Antoni}, {and} \bibinfo{person}{Thomas~W. Reps}.}
  \bibinfo{year}{2021}\natexlab{}.
\newblock \showarticletitle{Semantics-guided synthesis}.
\newblock \bibinfo{journal}{\emph{Proc. {ACM} Program. Lang.}}
  \bibinfo{volume}{5}, \bibinfo{number}{{POPL}} (\bibinfo{year}{2021}),
  \bibinfo{pages}{1--32}.
\newblock
\urldef\tempurl%
\url{https://doi.org/10.1145/3434311}
\showDOI{\tempurl}


\bibitem[Komuravelli et~al\mbox{.}(2016)]%
        {KomuravelliSpacer2016}
\bibfield{author}{\bibinfo{person}{Anvesh Komuravelli}, \bibinfo{person}{Arie
  Gurfinkel}, {and} \bibinfo{person}{Sagar Chaki}.}
  \bibinfo{year}{2016}\natexlab{}.
\newblock \showarticletitle{SMT-based model checking for recursive programs}.
\newblock \bibinfo{journal}{\emph{Formal Methods Syst. Des.}}
  \bibinfo{volume}{48}, \bibinfo{number}{3} (\bibinfo{year}{2016}),
  \bibinfo{pages}{175--205}.
\newblock
\urldef\tempurl%
\url{https://doi.org/10.1007/s10703-016-0249-4}
\showDOI{\tempurl}


\bibitem[Lee et~al\mbox{.}(2018)]%
        {lee-euphony-2018}
\bibfield{author}{\bibinfo{person}{Woosuk Lee}, \bibinfo{person}{Kihong Heo},
  \bibinfo{person}{Rajeev Alur}, {and} \bibinfo{person}{Mayur Naik}.}
  \bibinfo{year}{2018}\natexlab{}.
\newblock \showarticletitle{Accelerating search-based program synthesis using
  learned probabilistic models}. In \bibinfo{booktitle}{\emph{Proceedings of
  the 39th {ACM} {SIGPLAN} Conference on Programming Language Design and
  Implementation, {PLDI} 2018, Philadelphia, PA, USA, June 18-22, 2018}},
  \bibfield{editor}{\bibinfo{person}{Jeffrey~S. Foster} {and}
  \bibinfo{person}{Dan Grossman}} (Eds.). \bibinfo{publisher}{{ACM}},
  \bibinfo{pages}{436--449}.
\newblock
\urldef\tempurl%
\url{https://doi.org/10.1145/3192366.3192410}
\showDOI{\tempurl}


\bibitem[Oh et~al\mbox{.}(2012)]%
        {OhHLLY12}
\bibfield{author}{\bibinfo{person}{Hakjoo Oh}, \bibinfo{person}{Kihong Heo},
  \bibinfo{person}{Wonchan Lee}, \bibinfo{person}{Woosuk Lee}, {and}
  \bibinfo{person}{Kwangkeun Yi}.} \bibinfo{year}{2012}\natexlab{}.
\newblock \showarticletitle{Design and implementation of sparse global analyses
  for C-like languages}. In \bibinfo{booktitle}{\emph{{ACM} {SIGPLAN}
  Conference on Programming Language Design and Implementation, {PLDI} '12,
  Beijing, China - June 11 - 16, 2012}}. \bibinfo{publisher}{{ACM}},
  \bibinfo{pages}{229--238}.
\newblock
\urldef\tempurl%
\url{https://doi.org/10.1145/2254064.2254092}
\showDOI{\tempurl}


\bibitem[Ortega et~al\mbox{.}(2007)]%
        {OrtegaCA07}
\bibfield{author}{\bibinfo{person}{Alfonso Ortega}, \bibinfo{person}{Marina
  de~la Cruz}, {and} \bibinfo{person}{Manuel Alfonseca}.}
  \bibinfo{year}{2007}\natexlab{}.
\newblock \showarticletitle{Christiansen Grammar Evolution: Grammatical
  Evolution With Semantics}.
\newblock \bibinfo{journal}{\emph{{IEEE} Trans. Evol. Comput.}}
  \bibinfo{volume}{11}, \bibinfo{number}{1} (\bibinfo{year}{2007}),
  \bibinfo{pages}{77--90}.
\newblock
\urldef\tempurl%
\url{https://doi.org/10.1109/TEVC.2006.880327}
\showDOI{\tempurl}


\bibitem[Osera and Zdancewic(2015)]%
        {osera2015myth}
\bibfield{author}{\bibinfo{person}{Peter{-}Michael Osera} {and}
  \bibinfo{person}{Steve Zdancewic}.} \bibinfo{year}{2015}\natexlab{}.
\newblock \showarticletitle{Type-and-example-directed program synthesis}. In
  \bibinfo{booktitle}{\emph{Proceedings of the 36th {ACM} {SIGPLAN} Conference
  on Programming Language Design and Implementation, Portland, OR, USA, June
  15-17, 2015}}, \bibfield{editor}{\bibinfo{person}{David Grove} {and}
  \bibinfo{person}{Stephen~M. Blackburn}} (Eds.). \bibinfo{publisher}{{ACM}},
  \bibinfo{pages}{619--630}.
\newblock
\urldef\tempurl%
\url{https://doi.org/10.1145/2737924.2738007}
\showDOI{\tempurl}


\bibitem[Phothilimthana et~al\mbox{.}(2019)]%
        {PhothilimthanaE19}
\bibfield{author}{\bibinfo{person}{Phitchaya~Mangpo Phothilimthana},
  \bibinfo{person}{Archibald~Samuel Elliott}, \bibinfo{person}{An Wang},
  \bibinfo{person}{Abhinav Jangda}, \bibinfo{person}{Bastian Hagedorn},
  \bibinfo{person}{Henrik Barthels}, \bibinfo{person}{Samuel~J. Kaufman},
  \bibinfo{person}{Vinod Grover}, \bibinfo{person}{Emina Torlak}, {and}
  \bibinfo{person}{Rastislav Bod{\'{\i}}k}.} \bibinfo{year}{2019}\natexlab{}.
\newblock \showarticletitle{Swizzle Inventor: Data Movement Synthesis for {GPU}
  Kernels}. In \bibinfo{booktitle}{\emph{Proceedings of the Twenty-Fourth
  International Conference on Architectural Support for Programming Languages
  and Operating Systems, {ASPLOS} 2019, Providence, RI, USA, April 13-17,
  2019}}. \bibinfo{publisher}{{ACM}}, \bibinfo{pages}{65--78}.
\newblock
\urldef\tempurl%
\url{https://doi.org/10.1145/3297858.3304059}
\showDOI{\tempurl}


\bibitem[Polikarpova et~al\mbox{.}(2016)]%
        {Polikarpova2016Synquid}
\bibfield{author}{\bibinfo{person}{Nadia Polikarpova}, \bibinfo{person}{Ivan
  Kuraj}, {and} \bibinfo{person}{Armando Solar{-}Lezama}.}
  \bibinfo{year}{2016}\natexlab{}.
\newblock \showarticletitle{Program synthesis from polymorphic refinement
  types}. In \bibinfo{booktitle}{\emph{Proceedings of the 37th {ACM} {SIGPLAN}
  Conference on Programming Language Design and Implementation, {PLDI} 2016,
  Santa Barbara, CA, USA, June 13-17, 2016}},
  \bibfield{editor}{\bibinfo{person}{Chandra Krintz} {and}
  \bibinfo{person}{Emery~D. Berger}} (Eds.). \bibinfo{publisher}{{ACM}},
  \bibinfo{pages}{522--538}.
\newblock
\urldef\tempurl%
\url{https://doi.org/10.1145/2908080.2908093}
\showDOI{\tempurl}


\bibitem[Polozov and Gulwani(2015)]%
        {flashmeta}
\bibfield{author}{\bibinfo{person}{Oleksandr Polozov} {and}
  \bibinfo{person}{Sumit Gulwani}.} \bibinfo{year}{2015}\natexlab{}.
\newblock \showarticletitle{FlashMeta: A Framework for Inductive Program
  Synthesis}. In \bibinfo{booktitle}{\emph{Proceedings of the 2015 ACM SIGPLAN
  International Conference on Object-Oriented Programming, Systems, Languages,
  and Applications}} (Pittsburgh, PA, USA) \emph{(\bibinfo{series}{OOPSLA
  2015})}. \bibinfo{publisher}{Association for Computing Machinery},
  \bibinfo{address}{New York, NY, USA}, \bibinfo{pages}{107–126}.
\newblock
\showISBNx{9781450336895}
\urldef\tempurl%
\url{https://doi.org/10.1145/2814270.2814310}
\showDOI{\tempurl}


\bibitem[Raghothaman and Udupa(2014)]%
        {RaghothamanU14}
\bibfield{author}{\bibinfo{person}{Mukund Raghothaman} {and}
  \bibinfo{person}{Abhishek Udupa}.} \bibinfo{year}{2014}\natexlab{}.
\newblock \showarticletitle{Language to Specify Syntax-Guided Synthesis
  Problems}.
\newblock \bibinfo{journal}{\emph{CoRR}}  \bibinfo{volume}{abs/1405.5590}
  (\bibinfo{year}{2014}).
\newblock
\showeprint[arXiv]{1405.5590}
\urldef\tempurl%
\url{http://arxiv.org/abs/1405.5590}
\showURL{%
\tempurl}


\bibitem[Reback et~al\mbox{.}(2022)]%
        {pandas144}
\bibfield{author}{\bibinfo{person}{Jeff Reback},
  \bibinfo{person}{jbrockmendel}, \bibinfo{person}{Wes McKinney},
  \bibinfo{person}{Joris~Van den Bossche}, \bibinfo{person}{Matthew Roeschke},
  \bibinfo{person}{Tom Augspurger}, \bibinfo{person}{Simon Hawkins},
  \bibinfo{person}{Phillip Cloud}, \bibinfo{person}{gfyoung},
  \bibinfo{person}{Patrick Hoefler}, \bibinfo{person}{Sinhrks},
  \bibinfo{person}{Adam Klein}, \bibinfo{person}{Terji Petersen},
  \bibinfo{person}{Jeff Tratner}, \bibinfo{person}{Chang She},
  \bibinfo{person}{William Ayd}, \bibinfo{person}{Richard Shadrach},
  \bibinfo{person}{Shahar Naveh}, \bibinfo{person}{Marc Garcia},
  \bibinfo{person}{JHM Darbyshire}, \bibinfo{person}{Jeremy Schendel},
  \bibinfo{person}{Torsten Wörtwein}, \bibinfo{person}{Andy Hayden},
  \bibinfo{person}{Daniel Saxton}, \bibinfo{person}{Marco~Edward Gorelli},
  \bibinfo{person}{Fangchen Li}, \bibinfo{person}{Matthew Zeitlin},
  \bibinfo{person}{Vytautas Jancauskas}, \bibinfo{person}{Ali McMaster}, {and}
  \bibinfo{person}{Thomas Li}.} \bibinfo{year}{2022}\natexlab{}.
\newblock \bibinfo{booktitle}{\emph{pandas-dev/pandas: Pandas 1.4.4}}.
\newblock
\urldef\tempurl%
\url{https://doi.org/10.5281/zenodo.7037953}
\showDOI{\tempurl}


\bibitem[Reynolds et~al\mbox{.}(2015)]%
        {ReynoldsDKTB15}
\bibfield{author}{\bibinfo{person}{Andrew Reynolds}, \bibinfo{person}{Morgan
  Deters}, \bibinfo{person}{Viktor Kuncak}, \bibinfo{person}{Cesare Tinelli},
  {and} \bibinfo{person}{Clark~W. Barrett}.} \bibinfo{year}{2015}\natexlab{}.
\newblock \showarticletitle{Counterexample-Guided Quantifier Instantiation for
  Synthesis in {SMT}}. In \bibinfo{booktitle}{\emph{Computer Aided Verification
  - 27th International Conference, {CAV} 2015, San Francisco, CA, USA, July
  18-24, 2015, Proceedings, Part {II}}} \emph{(\bibinfo{series}{Lecture Notes
  in Computer Science}, Vol.~\bibinfo{volume}{9207})},
  \bibfield{editor}{\bibinfo{person}{Daniel Kroening} {and}
  \bibinfo{person}{Corina~S. Pasareanu}} (Eds.). \bibinfo{publisher}{Springer},
  \bibinfo{pages}{198--216}.
\newblock
\urldef\tempurl%
\url{https://doi.org/10.1007/978-3-319-21668-3\_12}
\showDOI{\tempurl}


\bibitem[Reynolds and Tinelli(2017)]%
        {reynoldscvc42017}
\bibfield{author}{\bibinfo{person}{Andrew Reynolds} {and}
  \bibinfo{person}{Cesare Tinelli}.} \bibinfo{year}{2017}\natexlab{}.
\newblock \showarticletitle{SyGuS Techniques in the Core of an {SMT} Solver}.
  In \bibinfo{booktitle}{\emph{Proceedings Sixth Workshop on Synthesis,
  SYNT@CAV 2017, Heidelberg, Germany, 22nd July 2017}}
  \emph{(\bibinfo{series}{{EPTCS}}, Vol.~\bibinfo{volume}{260})},
  \bibfield{editor}{\bibinfo{person}{Dana Fisman} {and} \bibinfo{person}{Swen
  Jacobs}} (Eds.). \bibinfo{pages}{81--96}.
\newblock
\urldef\tempurl%
\url{https://doi.org/10.4204/EPTCS.260.8}
\showDOI{\tempurl}


\bibitem[So and Oh(2017)]%
        {sosimpl2017}
\bibfield{author}{\bibinfo{person}{Sunbeom So} {and} \bibinfo{person}{Hakjoo
  Oh}.} \bibinfo{year}{2017}\natexlab{}.
\newblock \showarticletitle{Synthesizing Imperative Programs from Examples
  Guided by Static Analysis}. In \bibinfo{booktitle}{\emph{Static Analysis -
  24th International Symposium, {SAS} 2017, New York, NY, USA, August 30 -
  September 1, 2017, Proceedings}} \emph{(\bibinfo{series}{Lecture Notes in
  Computer Science}, Vol.~\bibinfo{volume}{10422})},
  \bibfield{editor}{\bibinfo{person}{Francesco Ranzato}} (Ed.).
  \bibinfo{publisher}{Springer}, \bibinfo{pages}{364--381}.
\newblock
\urldef\tempurl%
\url{https://doi.org/10.1007/978-3-319-66706-5\_18}
\showDOI{\tempurl}


\bibitem[Solar{-}Lezama(2013)]%
        {SolarLezamaSketch13}
\bibfield{author}{\bibinfo{person}{Armando Solar{-}Lezama}.}
  \bibinfo{year}{2013}\natexlab{}.
\newblock \showarticletitle{Program sketching}.
\newblock \bibinfo{journal}{\emph{Int. J. Softw. Tools Technol. Transf.}}
  \bibinfo{volume}{15}, \bibinfo{number}{5-6} (\bibinfo{year}{2013}),
  \bibinfo{pages}{475--495}.
\newblock
\urldef\tempurl%
\url{https://doi.org/10.1007/s10009-012-0249-7}
\showDOI{\tempurl}


\bibitem[Torlak and Bod{\'{\i}}k(2014)]%
        {TorlakRosette14}
\bibfield{author}{\bibinfo{person}{Emina Torlak} {and}
  \bibinfo{person}{Rastislav Bod{\'{\i}}k}.} \bibinfo{year}{2014}\natexlab{}.
\newblock \showarticletitle{A lightweight symbolic virtual machine for
  solver-aided host languages}. In \bibinfo{booktitle}{\emph{{ACM} {SIGPLAN}
  Conference on Programming Language Design and Implementation, {PLDI} '14,
  Edinburgh, United Kingdom - June 09 - 11, 2014}},
  \bibfield{editor}{\bibinfo{person}{Michael F.~P. O'Boyle} {and}
  \bibinfo{person}{Keshav Pingali}} (Eds.). \bibinfo{publisher}{{ACM}},
  \bibinfo{pages}{530--541}.
\newblock
\urldef\tempurl%
\url{https://doi.org/10.1145/2594291.2594340}
\showDOI{\tempurl}


\bibitem[Vechev et~al\mbox{.}(2010)]%
        {VechevYY10}
\bibfield{author}{\bibinfo{person}{Martin~T. Vechev}, \bibinfo{person}{Eran
  Yahav}, {and} \bibinfo{person}{Greta Yorsh}.}
  \bibinfo{year}{2010}\natexlab{}.
\newblock \showarticletitle{Abstraction-guided synthesis of synchronization}.
  In \bibinfo{booktitle}{\emph{Proceedings of the 37th {ACM} {SIGPLAN-SIGACT}
  Symposium on Principles of Programming Languages, {POPL} 2010, Madrid, Spain,
  January 17-23, 2010}}, \bibfield{editor}{\bibinfo{person}{Manuel~V.
  Hermenegildo} {and} \bibinfo{person}{Jens Palsberg}} (Eds.).
  \bibinfo{publisher}{{ACM}}, \bibinfo{pages}{327--338}.
\newblock
\urldef\tempurl%
\url{https://doi.org/10.1145/1706299.1706338}
\showDOI{\tempurl}


\bibitem[Wang et~al\mbox{.}(2017a)]%
        {Wang2017Scythe}
\bibfield{author}{\bibinfo{person}{Chenglong Wang}, \bibinfo{person}{Alvin
  Cheung}, {and} \bibinfo{person}{Rastislav Bod{\'{\i}}k}.}
  \bibinfo{year}{2017}\natexlab{a}.
\newblock \showarticletitle{Synthesizing highly expressive {SQL} queries from
  input-output examples}. In \bibinfo{booktitle}{\emph{Proceedings of the 38th
  {ACM} {SIGPLAN} Conference on Programming Language Design and Implementation,
  {PLDI} 2017, Barcelona, Spain, June 18-23, 2017}},
  \bibfield{editor}{\bibinfo{person}{Albert Cohen} {and}
  \bibinfo{person}{Martin~T. Vechev}} (Eds.). \bibinfo{publisher}{{ACM}},
  \bibinfo{pages}{452--466}.
\newblock
\urldef\tempurl%
\url{https://doi.org/10.1145/3062341.3062365}
\showDOI{\tempurl}


\bibitem[Wang et~al\mbox{.}(2017b)]%
        {syngar}
\bibfield{author}{\bibinfo{person}{Xinyu Wang}, \bibinfo{person}{Isil Dillig},
  {and} \bibinfo{person}{Rishabh Singh}.} \bibinfo{year}{2017}\natexlab{b}.
\newblock \showarticletitle{Program Synthesis Using Abstraction Refinement}.
\newblock \bibinfo{journal}{\emph{Proc. ACM Program. Lang.}}
  \bibinfo{volume}{2}, \bibinfo{number}{POPL}, Article \bibinfo{articleno}{63}
  (\bibinfo{date}{dec} \bibinfo{year}{2017}), \bibinfo{numpages}{30}~pages.
\newblock
\urldef\tempurl%
\url{https://doi.org/10.1145/3158151}
\showDOI{\tempurl}


\end{thebibliography}


\end{document}